\documentclass[manuscript,screen]{acmart}
\usepackage{multirow}
\usepackage{float}

\renewcommand\footnotetextcopyrightpermission[1]{}
\AtBeginDocument{%
  \providecommand\BibTeX{{%
    \normalfont B\kern-0.5em{\scshape i\kern-0.25em b}\kern-0.8em\TeX}}}

\copyrightyear{2025}
\acmYear{2025}
\setcopyright{cc}
\setcctype{by}
\acmConference[CHI '25]{CHI Conference on Human Factors in Computing Systems}{April 26-May 1, 2025}{Yokohama, Japan}
\acmBooktitle{CHI Conference on Human Factors in Computing Systems (CHI '25), April 26-May 1, 2025, Yokohama, Japan}\acmDOI{10.1145/3706598.3713920}
\acmISBN{979-8-4007-1394-1/25/04}

\begin{document}

\title[]{PANDA: Parkinson's Assistance and Notification Driving Aid}


\author{Tianyang Wen}
\authornote{Tianyang Wen and Jin Huang are also with the School of Computer Science and Technology, University of Chinese Academy of Sciences.}
\affiliation{%
  \institution{Institute of Software, Chinese Academy of Sciences} 
  \city{Beijing}
  \country{China} }
\email{wentianyang2023@iscas.ac.cn}

\author{Xucheng Zhang}
\affiliation{%
  \institution{Institute of Software, Chinese Academy of Sciences} 
  \city{Beijing}
  \country{China} }
\email{zhangxucheng@iscas.ac.cn}

\author{Zhirong Wan}
\affiliation{%
  \institution{Aerospace Center Hospital} 
  \city{Beijing}
  \country{China} }
\email{wanzhirong5587@126.com}

\author{Jing Zhao}
\affiliation{%
  \institution{Aerospace Center Hospital} 
  \city{Beijing}
  \country{China} }
\email{zhaoajing@foxmail.com}

\author{Yicheng Zhu}
\affiliation{%
  \institution{Peking Union Medical College Hospital} 
  \city{Beijing}
  \country{China} }
\email{zhuych910@163.com}

\author{Ning Su}
\affiliation{%
  \institution{Peking Union Medical College Hospital} 
  \city{Beijing}
  \country{China} }
\email{suning343@163.com}

\author{Xiaolan Peng}
\affiliation{%
  \institution{Institute of Software, Chinese Academy of Sciences} 
  \city{Beijing}
  \country{China} }
\email{xiaolan@iscas.ac.cn}

\author{Jin Huang}
\authornotemark[1]
\affiliation{%
  \institution{Institute of Software, Chinese Academy of Sciences}
  \city{Beijing}
  \country{China} }
\email{huangjin@iscas.ac.cn}

\author{Wei Sun}
\authornote{Corresponding authors}
\affiliation{%
  \institution{Institute of Software, Chinese Academy of Sciences}
  \city{Beijing}
  \country{China} }
\email{sunwei2017@iscas.ac.cn}

\author{Feng Tian}
\authornotemark[2]
\authornote{Feng Tian is also with the School of Artificial Intelligence, University of Chinese Academy of Sciences.}
\affiliation{%
  \institution{Institute of Software, Chinese Academy of Sciences} 
  \city{Beijing}
  \country{China} }
\email{tianfeng@iscas.ac.cn}

\author{Franklin Mingzhe Li}
\affiliation{%
  \institution{Carnegie Mellon University} 
  \city{Pittsburgh}
  \country{United States} }
\email{mingzhe2@cs.cmu.edu}

\renewcommand{\shortauthors}{Wen et al.}
\begin{abstract}

Parkinson's Disease (PD) significantly impacts driving abilities, often leading to early driving cessation or accidents due to reduced motor control and increasing reaction times. To diminish the impact of these symptoms, we developed PANDA (Parkinson's Assistance and Notification Driving Aid), a multi-modality real-time alert system designed to monitor driving patterns continuously and provide immediate alerts for irregular driving behaviors, enhancing driver safety of individuals with PD. The system was developed through a participatory design process with 9 people with PD and 13 non-PD individuals using a driving simulator, which allowed us to identify critical design characteristics and collect detailed data on driving behavior. A user study involving individuals with PD evaluated the effectiveness of PANDA, exploring optimal strategies for delivering alerts and ensuring they are timely and helpful. Our findings demonstrate that PANDA has the potential to enhance the driving safety of individuals with PD, offering a valuable tool for maintaining independence and confidence behind the wheel.

\end{abstract}



\begin{teaserfigure}
\centering
  \includegraphics[width=0.7\textwidth]{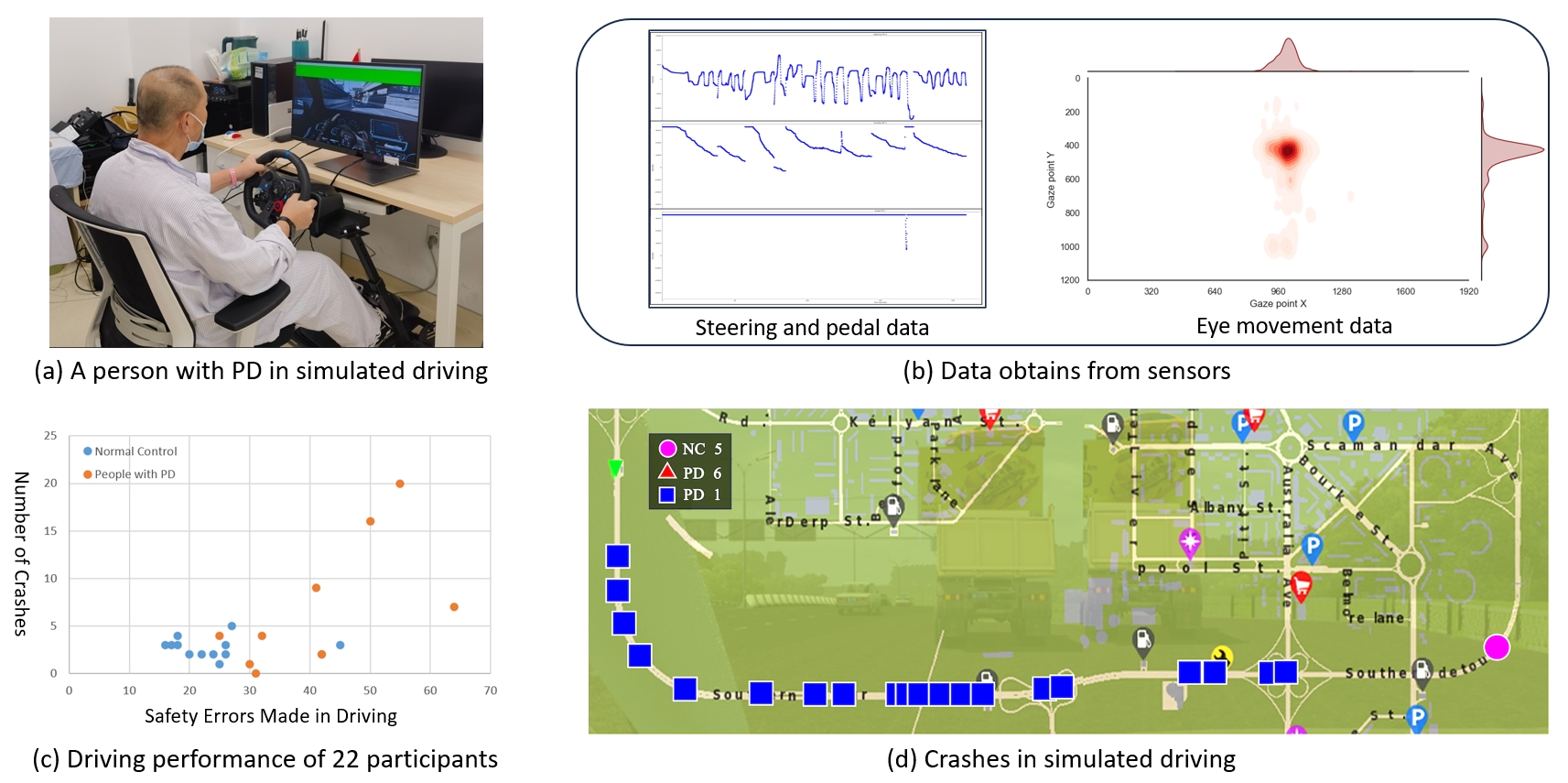}
  \caption{A simulated driving test involving 9 people with PD and 13 non-PD (normal control, NC) individuals showed considerable variation in driving abilities among people with PD. Notably, 5 out of 9 individuals with PD showed no decline in driving performance due to the disease, with their skills closing the average level of non-PD participants.}  
  \Description{This image contains four panels labeled (a), (b), (c), and (d), each depicting different aspects of a study on people with PD in a driving simulation.
  (a) A person with PD in simulated driving: This panel shows a person with PD seated in front of a driving simulator setup. The participant is holding a steering wheel positioned to simulate real driving within a controlled, experimental environment.
  (b) Data obtained from sensors: This panel presents data collected from various sensors during the simulation, including graphs of steering and pedal inputs on the left and a heatmap of eye movement data on the right, illustrating where the participant focuses during the simulation.
  (c) Driving performance of 22 participants: This scatter plot shows the driving performance of 22 participants, divided into two groups: Non-PD participants and people with PD. The x-axis represents "Safety Errors Made in Driving," while the y-axis shows the "Number of Crashes." Each point reflects individual performance, indicating that people with PD generally exhibit more errors and crashes than the control group.
  (d) Crashes in simulated driving: This map illustrates crash locations within the simulated driving environment for three specific participants: NC 5, PD 6, and PD 1. Each participant is represented by a unique symbol and color (pink circle for NC 5, red triangle for PD 6, and blue square for PD 1). The layout identifies where each participant experienced crashes during the virtual driving experiment.
  Overall, each panel contributes to the comprehensive analysis of how people with PD perform in a driving simulation, emphasizing crash locations, sensor data, and comparative driving performance.}
  \label{fig:teaser}
\end{teaserfigure}


\maketitle

\begin{figure*}[htbp!]
  \centering
  \includegraphics[width=0.5\textwidth]{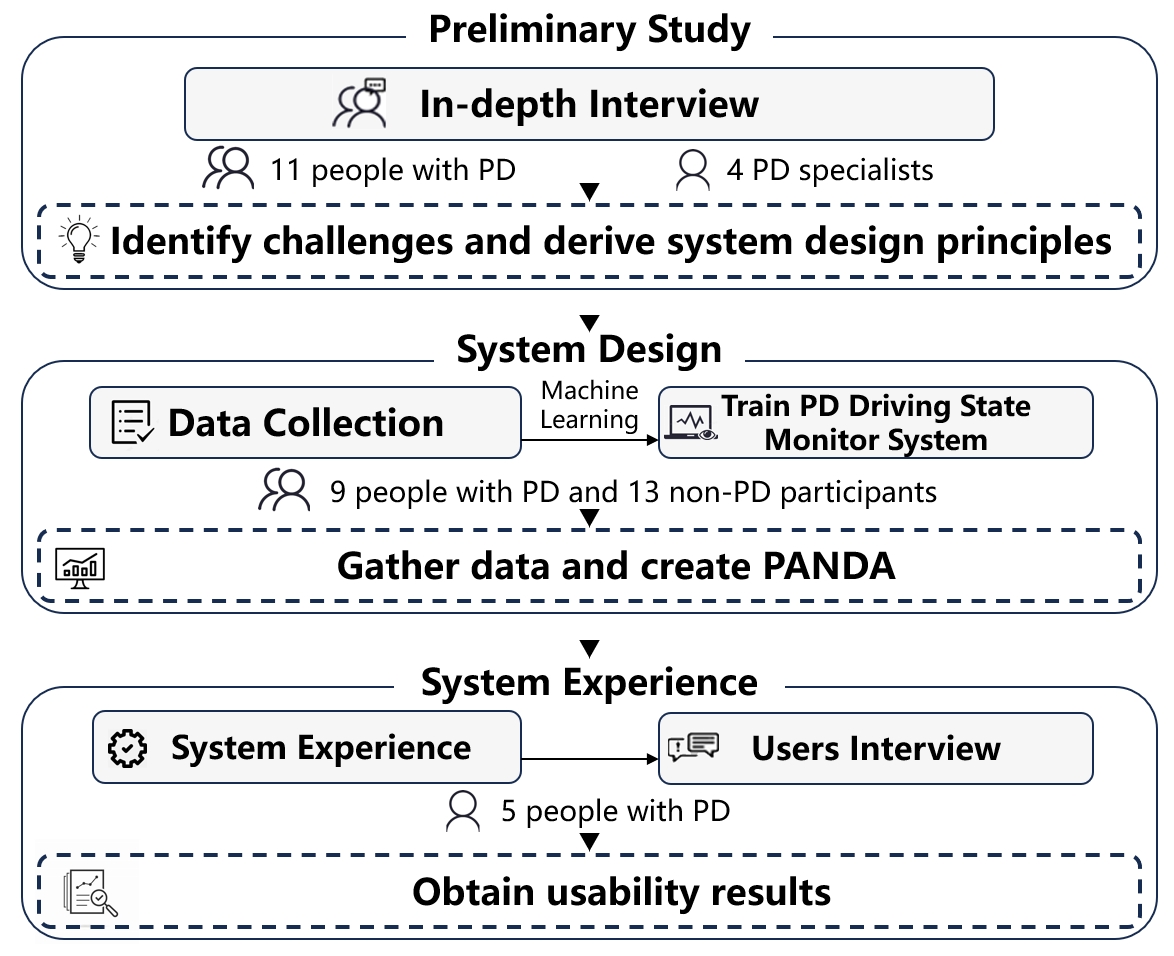}
  \caption{Our study employs a user-centered, three-step design process: In the preliminary study, we gathered challenges and system design principles through interviews. During the system design stage, we collected user operation data and used it to train a PD driving state monitor system. In the user study phase, we obtained feedback on the monitor system with various alerts and derived effective alert strategies for drivers with PD.}
  \Description{The diagram illustrates the research process in three main phases: Preliminary Study, System Design, and System Experience. Each phase includes specific activities, participant details, and goals, with arrows showing the flow between steps.
  Preliminary Study
  In-depth Interview: This stage involved interviewing 11 people with PD and 4 PD specialists. Objective: The goal was to identify challenges faced by people with PD and to derive system design principles based on these insights.
  System Design
  Data Collection: Data was gathered from 9 people with PD and 13 non-PD participants. Machine Learning Application: This data was used to train the "PD Driving State Monitor System," a monitoring system designed to track the driving states of people with PD. Outcome: Data collected was processed to create a system referred to as "PANDA."
  System Experience
  System Experience: This phase tested the system with 5 people with PD. Users Interview: After using the system, people with PD participated in interviews to provide feedback. 
  Objective: The goal was to obtain usability results.
  Each phase builds on the previous one, with the aim of developing a system that is responsive to the unique challenges of people with PD and supports their driving experience through monitoring and feedback.}
  \label{fig:methods.}
\end{figure*}

\section{INTRODUCTION}







Parkinson's Disease (PD) is a progressive neurological disorder that affects about 1.7\% of individuals over 65, with prevalence increasing with age \cite{zhang2005parkinson, stamatelos2024parkinson}. This equates to millions of people globally, many of whom are active drivers \cite{stamatelos2024parkinson}. PD impairs motor functions, cognitive abilities, and visual processing, all of which are critical for safe driving \cite{ranchet2020driving}. Simulated and real-world driving studies have revealed significant challenges faced by drivers with PD, particularly those with attentional or motor impairments \cite{thompson2018driving, dubois1996cognitive, ranchet2020driving}. These challenges lead to delayed or incorrect operations, posing risks not only to individuals with PD but also to public safety \cite{thompson2018driving, dubois1996cognitive, singh2006parkinson}. With the global aging population increasing rapidly, addressing driving challenges in PD is not just a public health issue but a societal imperative, as unsafe driving due to PD-related impairments impacts road safety, healthcare costs, and individual quality of life \cite{uc2009road, nishihori2015effect}. Additionally, premature cessation of driving can lead to social isolation and psychological issues, such as depression \cite{eichel2022neuropsychiatric}.

To address the risks associated with driving while managing PD symptoms, traditional driving evaluations are recommended every 6–12 months \cite{stamatelos2024parkinson}. These assessments typically involve multi-disciplinary teams conducting practical driving tests. However, these episodic evaluations are inadequate for capturing the day-to-day variability in driving ability caused by PD's fluctuating symptoms \cite{jankovic2005motor}. Furthermore, explicit driving assessment methods can add cognitive load, causing stress that may impair performance \cite{kihlstrom2004implicit}. Compounding these issues, people with PD and their doctors often overestimate driving abilities \cite{heikkila1998decreased}, leading to potential underestimation of risks. Additionally, the lack of specific driving regulations for individuals with PD in many countries contributes to reluctance to seek assessments due to fear of unfavorable outcomes \cite{alvarez2016parkinson}. Despite the publication of several guidelines, neurologists face challenges in providing evidence-based consultation about driving fitness, as the procedures and their roles in evaluations remain unclear \cite{stamatelos2024parkinson}.

The problem is inherently complex due to the dynamic nature of PD symptoms, which can be influenced by factors such as disease progression, mood, medication effects, and stress levels. This variability complicates the development of reliable, real-time systems to monitor driving ability. Traditional approaches often fail to provide continuous and adaptive feedback, leaving a critical gap in addressing real-world driving scenarios. Additionally, explicit evaluations conducted in controlled settings may not reflect actual driving conditions, and the stress induced by such evaluations can further skew results \cite{heikkila1998decreased}. Moreover, the fear of losing driving privileges discourages individuals from participating in assessments, limiting the effectiveness of current methods in ensuring public safety and individual well-being.

In light of these challenges, we explored the following research questions: 

\begin{itemize}
    \item What specific challenges do individuals with PD face while driving?
    \item What design principles should an effective driving assistance system incorporate?
    \item What real-time alert strategies are most effective in enhancing safety and user confidence?
\end{itemize}

To address these questions, we present PANDA (Parkinson's Assistance and Notification Driving Aid), a system that leverages data from eye-tracking, steering wheel, and pedal sensors to detect irregular driving patterns and provide timely alerts. Recognizing the increasing prevalence of sensor integration in modern vehicles, PANDA offers a practical and scalable solution for real-time and continuous assessment. Our approach is informed by three steps: 1) a preliminary interview with 11 individuals with PD and 4 PD specialists to identify key design principles, 2) simulated driving tasks involving 9 individuals with PD and 13 non-PD participants to collect data for model training, establish baseline behavior and refine alert strategies, and 3) a user study validating PANDA's usability with 5 individuals with PD. By addressing the limitations of traditional evaluations and integrating real-time feedback, PANDA aims to empower individuals with PD to drive safely and confidently in daily scenarios. The contributions of this paper are as follows:
\begin{itemize}
    \item We conducted a preliminary study with both 11 individuals with PD and 4 PD specialists to identify existing practices, perceptions, and challenges of people with PD regarding driving;
    \item We collected a multi-modality driving dataset from 22 participants, including both people with PD and non-PD individuals in simulated driving. To accelerate the development in this field, we provide this dataset on Google Drive\footnote{\url{https://drive.google.com/drive/folders/1bZSsoktWF9d_0AQhjSIrQYdPnwsaQPni?usp=drive_link}};
    \item We developed a novel machine learning model capable of detecting and distinguishing driving behaviors between people with PD and non-PD individuals in real-time;
    \item We designed a technology probe (PANDA) using our model and identified novel visual alert preferences, audio alert preferences, and design implications for driving monitoring and alerting systems for people with PD;
    \item We evaluated PANDA to understand the experiences and design implications of a real-time alert system with 5 people with PD.
\end{itemize}


\begin{table*}[h!]
\centering
\resizebox{\textwidth}{!}{
\begin{tabular}{c|cccccccccc}
\hline
\textbf{PID} & \textbf{Age} & \textbf{Gender}  & \textbf{Disease Duration} & \textbf{HY Scale} & \textbf{UPDRS-III} & \textbf{Education Yrs} & \textbf{Holding Driver's License Yrs} &\textbf{Driving Status} &\textbf{Participated EX \#}\\
\hline\hline
P1 & 62 & M & 2 years & 3 & 32 & 10 & 27 & stop driving 2 years & 1 and 2\\
P2 & 66 & M & 11 years & 2.5 & 32 & 14 & 23 & stop driving 17 years &  1 and 2 and 3\\
P3 & 53 & M & 6.5 years & 2.5 & 27 & 24 & 22 & 1-2 times a month & 1 and 2\\
P4 & 63 & M & 12 years & 1 & 17 & 14 & 31 & everyday & 1 and 2 and 3\\
P5 & 55 & M & 3 years & 1 & 10 & 16 & 27 & everyday & 1 and 2 and 3\\
P6 & 61 & M & 5 years & 2.5 & 42 & 13 & 33 & 3 times a week & 1 and 2 and 3\\
P7 & 52 & F & 1 years & 1.5 & 25 & 18 & 16 & 2 times a week & 1 and 2 and 3\\
P8 & 64 & M & 7 years & 2.5 & 23 & 11 & 44 & 5-6 times a month & 1 and 2\\
P9 & 65 & M & 11 years & 2.5 & 24 & 14 & 14 & stop driving 2 months & 1 and 2\\
P10 & 60 & M & 8 years & 2 & 23 & 11 & 25 & stop driving 2 years & 1\\
P11 & 61 & M & 10 years & 2 & 36 & 12 & 21 & stop driving 5 years & 1\\

\hline
\end{tabular}
}
\captionsetup{position=bottom} 
\caption{Demographic information of people with PD in our study. Participation is recorded in the last column. Experiment 1 for preliminary study, 2 for data collection, and 3 for user study.
We present the severity of PD in the form of Hoehn and Yahr Scale (HY Scale) \cite{goetz2004movement} and Unified Parkinson's Disease Rating Scale part three (UPDRS-III) \cite{movement2003unified}. The HY Scale is the most commonly used and widely accepted tool worldwide for describing the severity of PD, with scores ranging from 1 (mild) to 5 (severe) \cite{goetz2004movement}. Similarly, the UPDRS is a widely used clinical rating scale for PD, with its Part III (UPDRS-III) specifically designed for motor examination, where higher scores indicate greater motor impairment \cite{goetz2008movement}.} 

\Description{
The table includes data from 11 people with PD, organized as follows:
1. PID: Participant ID (ranging from 1 to 11).
2. Age: The age of each participant.
3. Gender: The gender of each participant (male or female).
4. Disease Duration: The number of years each participant has been living with PD.
5. HY Scale: The Hoehn and Yahr scale score was used to assess the severity of PD for each participant.
6. UPDRS-III: Unified Parkinson's Disease Rating Scale part III score, indicating motor function assessment for each participant.
7. Education Yrs: The number of years of formal education completed by each participant.
8. Holding Driver's License Yrs: The number of years each participant has held a driver's license.
9. Driving Status: Whether the participant is currently able to drive.
10. Participated EX #: The number of exercise sessions or activities each participant has participated in.
Each of the 11 rows in the table corresponds to a specific participant, showing their details across all the categories listed above.
}
\label{tab:patient_info}
\end{table*}

\section{RELATED WORK}
In this section, we review prior research on the impact of Parkinson's Disease (PD) on daily activities, its effects on driving performance, and advancements in driving monitoring and alerting systems.

\subsection{Parkinson's Disease in Daily Activity} 

Parkinson's disease (PD) is marked by motor symptoms such as tremors, rigidity, bradykinesia/akinesia, and postural instability \cite{balestrino2020parkinson}, all of which significantly affect daily activities and quality of life \cite{hariz2011activities,li2022freedom,tian2021natural,wang2021white}. Efforts to improve the lives of individuals with PD have resulted in various assistive technologies. For instance, specialized spoons designed to remain upright help mitigate tremors during eating \cite{al2022smart}, while wearable devices employing visual, auditory, or tactile cues assist with walking challenges \cite{bachlin2010wearable, canning2020virtual}. Driving, a vital aspect of independence and quality of life \cite{patterson2019measuring}, has also been extensively studied, but research primarily focuses on assessing driving impairments rather than developing tools to assist people with PD in driving safely. This represents an unmet need for targeted innovations in driving assistance for people with PD.


Real-time driving data is closely linked to driving status, and its crucial role in classifying driving conditions, such as fatigued driving and distracted driving, has been demonstrated in various tasks \cite{sikander2018driver, qi2020distracted}. However, doctors cannot access patients' driving data during the clinical stage, and the relationship between clinical data and driving ability remains widely debated \cite{crizzle2012parkinson}, making it difficult for doctors to accurately assess a patient's driving capability. We believe that obtaining and analyzing real-time driving data from individuals with PD, in conjunction with clinical data, can help doctors better assess the patient's driving status, thereby enabling individuals with PD to drive more independently and confidently.


\subsection{Effects of Parkinson's Disease on Driving}
PD is a non-lethal but irreversible neurodegenerative disease that presents with bradykinesia combined with either rest tremor, rigidity, or both \cite{bloem2021parkinson}.
PD significantly impairs an individual's driving performance, as demonstrated by numerous clinical studies. Research indicates that people with PD exhibit compromised movement, cognitive, and visual functions, resulting in irregular driving patterns such as slower reaction times, reduced speed, increased speed variability, and greater variability in lateral lane position \cite{weil2016visual,rodnitzky2013visual,farashi2021analysis,makhoul2023driving,ranchet2020driving}. These irregularities contribute to a 6.16-fold increase in on-the-road test failures and a 2.63-fold increase in simulator crashes compared to non-PD people \cite{thompson2018driving}. Moreover, driving performance is further diminished by increased daytime sleepiness and medication-related sleep attacks \cite{thompson2018driving,falkenstein2020age}. 
Consequently, periodic evaluations of driving fitness for people with PD are necessary, yet no uniform legal criteria exist for this purpose \cite{ranchet2020driving}. A recent study \cite{stamatelos2024parkinson} proposed a driving fitness evaluation procedure involving multidisciplinary doctor participation and practical driving assessments. Existing research primarily focuses on statistical analyses of PD's impact on driving performance and its correlation with neuropsychological test results \cite{klimkeit2009driving}. In addition, people with PD also have unique emotional needs and varying reactions to user interfaces under different conditions \cite{peng2025olderadults, zheng2021scenario}. To our knowledge, our study is the first to analyze real-time driving patterns in people with PD and develop a real-time monitoring system for their assessment.

\subsection{Monitoring and Alerting Systems for Driving}
The increasing frequency of serious car crashes has been linked to factors such as driver drowsiness, distraction, excessive mental workload, extreme emotions, and alcohol consumption, all of which significantly impair driving performance \cite{kang2013various, halin2021survey, koch2023leveraging, hu2013negative}. To address these risks, advanced sensing systems have been developed to monitor driving behavior and detect irregularities that may lead to accidents. These systems rely on diverse data sources, including vehicle status parameters such as steering wheel input, pedal activity, speed, and lane-keeping data, accessed via the CAN bus \cite{doudou2020driver, khan2019comprehensive,koesdwiady2016recent}. Additionally, driver behavior metrics are captured using in-vehicle cameras, microphones, and seat-embedded sensors, while external environmental data is collected through vehicle-mounted cameras \cite{chhabra2017survey, melnicuk2016towards,luo2020vr}. The acquired data is analyzed using rule-based models, mathematical frameworks, or machine learning techniques to assess driving states and detect irregular behaviors \cite{sikander2018driver,luo2025exploring}.

When irregular driving patterns are identified, alerts are typically delivered to the driver through visual, auditory, or tactile cues. These notifications are displayed on head-up displays (HUDs), dashboards, or central control screens \cite{winkler2018warn, zhang2020influence, janes2013effective}, ensuring timely responses to prevent potential accidents \cite{xu2019pilot, xu2022auditory, rohl2016evaluating}. While existing systems focus on general driving populations, individuals with PD may exhibit unique driving patterns as their condition progresses. 

This work, inspired by current driver state detection technologies, aims to explore the design of a tailored detection system that accommodates the specific needs and challenges of drivers with PD.

\section{PRELIMINARY STUDY}

In this section, we present interviews with 11 people with PD and 4 PD specialists. 
Our goal was to identify existing practices, perceptions, and challenges of people with PD regarding driving, in order to provide guidance for design decisions in the development of PANDA.

\subsection{Interview of People with PD}
\subsubsection{Participants}
\label{interview participants}
We conducted semi-structured interviews with 11 people with PD (Table \ref{tab:patient_info}). Among the 11 participants (10 male, 1 female), they have an average age of 60.2, an average duration of 7.0 years of PD, and an average year of 25.7 in driving. At the current stage, six of them still drive occasionally, and five of them stopped driving because of PD. They have an average Hoehn and Yahr Scale (HY Scale) of 2.1 and an average of Unified Parkinson's Disease Rating Scale, Part III (UPDRS-III) of 26.5 \cite{movement2003unified}, where HY Scale is the most commonly used and widely accepted tool worldwide for describing the severity of PD, with scores ranging from 1 (mild) to 5 (severe) \cite{goetz2004movement}, and the UPDRS is a widely used clinical rating scale for PD, with its Part III (UPDRS-III) specifically designed for motor examination, where higher scores indicate greater motor impairment \cite{goetz2008movement}. Participants in this study were compensated in local currency which is equivalent to \$40 USD. The recruitment and study procedure was approved by the local hospital's Institutional Review Board (IRB). The interview took approximately 20 minutes.
\subsubsection{Procedure}
The questions began with inquiries about changes in driving ability since the onset of the illness, their current driving status, and whether they had sought any assistance. For those who had stopped driving, we explored the reasons for cessation, the impact of quitting, and their willingness to drive again. Next, we detailed current methods for assessing driving in people with PD and asked participants to evaluate their feasibility based on their personal experiences, explaining their reasons. Finally, we presented our study's objectives, shared potential design ideas, and invited participants to contribute to the design principles and offer their suggestions.

\begin{table*}[!htbp]
\centering

\begin{tabular}{c|ccccccccc}
\hline
\textbf{PID} & \textbf{Gender}  & \textbf{Specialty} &\textbf{Years of Experience} & \textbf{Age}\\
\hline\hline
P1 & M & Specialist in Neurology & 21 & 46\\
P2 & F & Specialist in Neurology & 11 & 38\\
P3 & F & Specialist in Neurology & 30 & 52\\
P4 & F & Specialist in Neurology & 8 & 38\\

\hline
\end{tabular}

\captionsetup{position=bottom} 
\caption{Demographic information of PD specialists in preliminary study} 
\Description{
The table contains data for four PD specialists, with the following details:
1. PID: Participant ID (1 to 4).
2. Gender: The gender of each participant (male or female).
3. Specialty: The professional specialty or field of expertise of each participant.
4. Years of Experience: The number of years each participant has been working in their specialty.
5. Age: The age of each participant.
Each of the 4 rows in the table corresponds to a specific participant, showing their details across all the categories listed above.
}
\label{tab:specilist_info}
\end{table*}

\subsection{Interview of PD Specialists}
To understand the neural mechanisms underlying PD and the challenges people with PD face while driving, we conducted interviews with PD specialists. 
\subsubsection{Participants}
We recruited four PD Specialists (3 Female, 1 Male) with an average age of 43.5 from local hospitals, all of whom specialize in neurology (Table \ref{tab:specilist_info}). They had an average experience in PD for 18.5 years.

\subsubsection{Procedure}
Our preliminary interview centered on the following key areas:
1) The symptoms most affecting the driving abilities of individuals with PD, their impact on performance, and specific road situations or emergency events that present significant challenges.
2) The common responses specialists provide to patients' inquiries about driving and the advice given.
3) The effects of various medications on driving performance, including the duration of their impact and potential side effects that could influence driving safety.
4) The current methods used for assessing driving performance in people with PD, from the perspective of PD specialists.
5) The effectiveness of our current design principles and any additional design factors specialists would suggest.
6) The most effective notification strategies and optimal alert buffer time for people with PD. 
The interview took around 15 minutes.

\subsection{Data Analysis}
We recorded the audio during the interviews and used handwritten notes to record important information.
After conducting the interviews, the findings were analyzed using thematic analysis \cite{braun2006using}.

Firstly, all audio recordings were transcribed into text and cross-checked against handwritten notes to serve as the data input. Secondly, two researchers independently employed thematic analysis to identify, analyze, and interpret themes in the data. Any conflicts in theme identification were resolved through discussion and consensus. 
In this study, we use Cohen's Kappa coefficient~\cite{mchugh2012interrater} to quantify the degree of agreement between two researchers. The coefficient ranges from -1 (indicating perfect disagreement) to +1 (indicating perfect agreement), with values closer to +1 reflecting stronger reliability.
The analysis of interviews with people with PD focused on three themes: demand, factors impacting driving, and challenges (as shown in Table~\ref{tab:thematic analysis pre patients} in Appendix). 
The Cohen's Kappa coefficient for inter-rater reliability between the two researchers was 0.727. The analysis of interviews with PD specialists focused on three themes: current solutions, factors impacting driving, and challenges (as shown in Table~\ref{tab:thematic analysis pre specialists} in Appendix). The Cohen's Kappa coefficient for inter-rater reliability between the two researchers was 0.794.


\begin{figure*}[!htbp]
  \centering
  \resizebox{\textwidth}{!}{\includegraphics{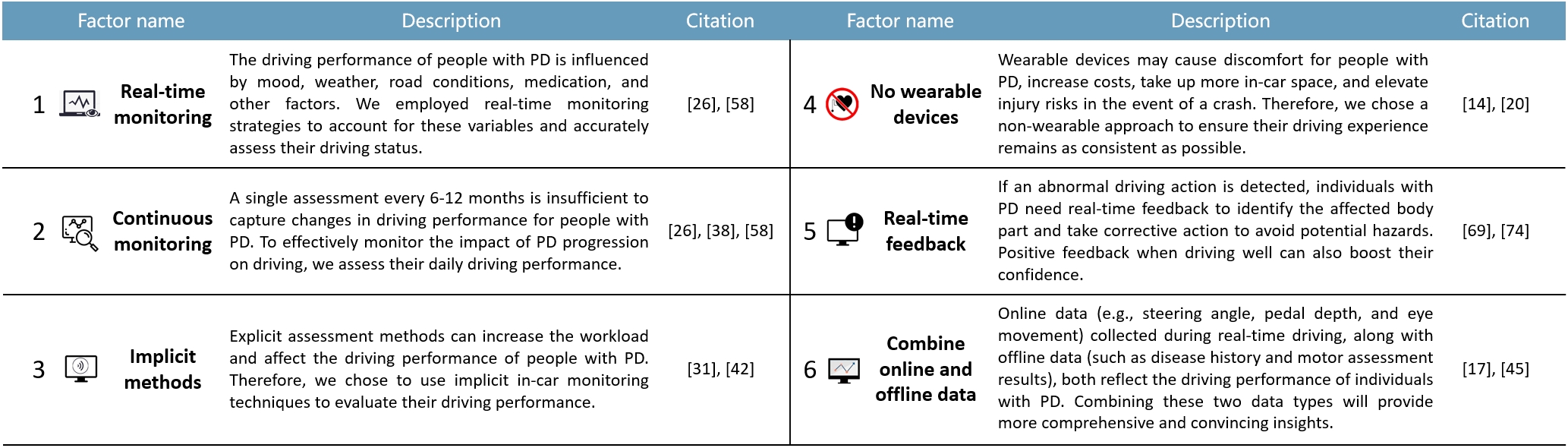}} 
  \caption{From interviews of people with PD and PD specialists and existing literature, we identify six critical design principles. This figure provides an overview of these principles. 
  }
  \Description{The table presents six factors to consider for monitoring the driving performance of people with PD, each with a description explaining the factor's importance and corresponding citations.
  1. Real-time monitoring. Description: The driving performance of people with PD is influenced by mood, weather, road conditions, medication, and other factors. We employed real-time monitoring strategies to account for these variables and accurately assess their driving status.
  2. Continuous monitoring. Description: A single assessment every 6-12 months is insufficient to capture changes in driving performance for people with PD. To effectively monitor the impact of PD progression on driving, we assess their daily driving performance.
  3. Implicit methods. Description: Explicit assessment methods can increase the workload and affect the driving performance of people with PD. Therefore, we chose to use implicit in-car monitoring techniques to evaluate their driving performance.
  4. No wearable devices. Description: Wearable devices may cause discomfort for people with PD, increase costs, take up more in-car space, and elevate injury risks in the event of a crash. Therefore, we chose a non-wearable approach to ensure their driving experience remains as consistent as possible.
  5. Real-time feedback. Description: If an abnormal driving action is detected, individuals with PD need real-time feedback to identify the affected body part and take corrective action to avoid potential hazards. Positive feedback when driving well can also boost their confidence.
  6. Combine online and offline data. Description: Online data (e.g., steering angle, pedal depth, and eye movement) collected during real-time driving, along with offline data (such as disease history and motor assessment results), both reflect the driving performance of individuals with PD. Combining these two data types will provide more comprehensive and convincing insights.}
  \label{fig:design principles.}
\end{figure*}

\subsection{Findings}
After synthesizing the interview results and incorporating knowledge from previous work, the preliminary study identified several challenges and key design principles for a PD driving assistance system (Figure \ref{fig:design principles.}).
\subsubsection{The ability to drive is crucial for individuals with PD, yet there are currently no effective objective assessment methods available.}

Maintaining the ability to drive is crucial for preserving independence and plays a significant role in the self-esteem of people with PD. Many individuals with PD express a strong desire to continue driving, as it significantly enhances their convenience and quality of life. For instance, P10, who experiences Freezing of Gait (FoG), a PD symptom characterized by brief episodes of inability to move or very short steps, typically occurring when starting to walk or turning \cite{nutt2011freezing}, stated, "\textit{I need to go to the market 3-4 times a week. Walking there is very challenging for me, but driving makes it much easier to get there.}" 
Surprisingly, despite PD impairing his foot movement and causing difficulty with walking, he is still able to operate the pedals as if he were unaffected. This underscores how driving remains an essential aspect of daily life for many people with PD. However, from the perspective of PD specialists, the primary concern is helping patients maintain essential functions such as walking steadily and standing up securely. Most specialists do not consider driving ability a key aspect of evaluation in clinical settings. Regarding policy issues, there are no uniform legal criteria to guide people with PD and doctors on assessing driving fitness \cite{ranchet2020driving}.

\subsubsection{The driving status of people with PD can change rapidly and is influenced by a variety of factors. Real-time and continuous driving assessment is urgently needed.}
\label{sec:driving status}

Everyone's driving status is affected by road conditions, weather, and mood.
While people with PD also experience these influencing factors, they are also affected by additional issues. 
Long-term levodopa therapy can lead to the 'on-off' phenomenon, characterized by fluctuations in the severity of PD symptoms \cite{marsden1976off}. Additionally, medication side effects often cause drowsiness in people with PD \cite{friedman2007fatigue}, further impacting their driving ability. P9 requested to discontinue the experiment in the afternoon, stating, "\textit{I always feel sleepy after noon, and my condition is better in the morning. I'll return to complete the experiment another morning.}" P3 reported feeling very unwell after the test, mentioning, "\textit{The effect of my medication has worn off; I need to take a bit more.}"
PD can cause photophobia \cite{hori2008pupillary}, making it difficult for some patients to drive safely when the sun is strong. 
P1 drove with his eyes slightly open and said, "\textit{It's hard for me to look up when the light is strong since I developed PD.}"
Given these real-time influencing factors, a single assessment cannot accurately represent the daily driving conditions of people with PD. Therefore, real-time and continuous assessment is urgently needed (Figure~\ref{fig:design principles.}).

\subsubsection{Given the unique challenges faced by people with PD, designing an irregular driving alert system tailored to their needs requires careful exploration and thoughtful consideration.}
\label{driving alert}
PD can impair both cognitive and visual functions in patients \cite{aarsland2021parkinson,weil2016visual}. Designing visual and auditory alerts that effectively capture attention without disrupting normal driving is still an area of active research (Figure~\ref{fig:design principles.}). P5 shared, “\textit{Since my diagnosis, I've noticed that my reactions have slowed and handling multiple tasks simultaneously has become more difficult.}”
Our interviews revealed that people with PD often display similar driving patterns due to shared symptoms, such as delayed responses, unintended lane deviations, shaky steering from hand tremors, and weakened pedal control (Figure~\ref{fig:design principles.}). These findings align with existing literature \cite{madeley1990,wood2005}. We speculate that alerting methods may need to differ across various driving scenarios to ensure that the system remains sensitive during critical events while avoiding excessive notifications (Figure~\ref{fig:design principles.}). Identifying which specific situations require the most urgent alerts remains a subject for further research.
Moreover, designing an alert system for people with PD must address privacy and social considerations. The disease often brings about psychological challenges \cite{garlovsky2016psychological}, making it crucial for any system to support drivers while preserving their dignity. This includes offering features that prevent the disclosure of sensitive alerts to other passengers, thereby maintaining both their self-esteem and driving independence.
Regarding the optimal time buffer for deploying system alerts, after a detailed explanation of the design rationale for the buffer, PD specialist P4 noted: "A 10-second buffer is optimal. If the buffer duration is too long, it may delay the best opportunity to avoid a hazard, while a buffer that is too short requires excessively frequent validation."

\section{EXPERIMENTAL SETUP AND DATA COLLECTION}
\label{data collection}

Drawing on insights from the preliminary study, we identified the critical importance and inherent challenges of driving for individuals with PD. This understanding informed the development of a real-time, continuous monitoring system. To support this, we created a simulated driving environment and designed experimental paradigms that closely mirror real-world driving scenarios. We collected driving behavior data from both individuals with and without PD, and used this data to train a machine learning model capable of detecting irregular driving patterns associated with PD.

\begin{table*}[!htbp]
\centering

\begin{tabular}{cccc}
\hline
\textbf{ } & \textbf{Individuals with PD}& \textbf{Non-PD participants}& \textbf{Sig}\\
\hline
n (total n = 22) & 9 & 13 & n/a\\
\hline
\textbf{Demographics}\\
\hline
Women \# (\%)  &  1 (11\%) & 5 (38\%) &n/a\\
Men \# (\%)  &  8 (89\%) & 8 (62\%) & n/a\\
Age yrs (std)  &  60.1 (5.0) & 61.38 (5.8) & p = 0.71\\
Education yrs (std)  & 14.9 (3.9) & 14.1 (2.3) & p = 1.0\\
Obtain Driver' License yrs (std)  & 26.3 (8.6) & 24.6 (4.8) & p = 0.76\\
\hline
\textbf{Clinical Characteristics}\\
\hline
Disease Onset yrs (std) & 6.5 (3.9)& 0 (0) & p < 0.001\\
UPDRS Part III \cite{movement2003unified} avg (std) &25.8 (8.7)& 0 (0) & p < 0.001\\
Hoehn-Yahr \cite{bhidayasiri2012parkinson} avg (std) &2.1 (0.7)& 0 (0) & p < 0.001\\

\hline
\end{tabular}

\captionsetup{position=bottom} 
\caption{Summary of the demographic and clinical information of the 22 participants in this study.} 
\Description{
    The table compares two groups, individuals with PD and non-PD participants, across demographic and clinical characteristics. The data is as follows:    
    Demographics:
    Total Participants: 22 (9 individuals with PD, 13 non-PD).
    Women: 1 (11\%) with PD, 5 (38\%) non-PD.
    Men: 8 (89\%) with PD, 8 (62\%) non-PD.
    Age (average ± standard deviation):
    PD: 60.1 ± 5.0 years
    Non-PD: 61.38 ± 5.8 years
    Statistical significance (p-value) = 0.71.
    Education (average ± standard deviation):
    PD: 14.9 ± 3.9 years
    Non-PD: 14.1 ± 2.3 years
    p-value = 1.0.
    Years of holding a driver's license (average ± standard deviation):
    PD: 26.3 ± 8.6 years
    Non-PD: 24.6 ± 4.8 years
    p-value = 0.76.
    Clinical Characteristics:
    Disease Onset (average ± standard deviation):
    PD: 6.5 ± 3.9 years
    Non-PD: 0
    p-value < 0.001.
    UPDRS Part III (average ± standard deviation):
    PD: 25.8 ± 8.7
    Non-PD: 0 
    p-value < 0.001.
    Hoehn-Yahr Scale (average ± standard deviation):
    PD: 2.1 ± 0.7
    Non-PD: 0
    p-value < 0.001.
}
\label{tab:participant}
\end{table*}

\subsection{Driving System Setup}
This experiment set up a simulated driving system including an eye-tracker (Tobii Pro X3-120), steering wheel (Logitech G29), and pedals (accelerator, brake, and paddle shifters), as shown in Figure \ref{fig:teaser}.
The eye-tracker captured the sight changes during driving, such as eye movement and fixation state, with a 120Hz sample rate. These changes may indicate some reactions or human perceptions while driving. Before each experimental scenario, the eye tracker was calibrated. 
The steering wheel and pedals recorded the rotation angle and pedaling amplitude, providing detailed motion data of hand and foot movements, with a sampling rate of 20Hz. 

The simulated driving environments, along with the tasks, were selected in a software which is specially designed for driving test training purposes named City Car Driving\footnote{https://citycardriving.com/}, using a standard left-hand driving automatic sedan. This software vividly renders many real-life driving elements, including the dashboard, traffic lights, traffic signs, pedestrians, other vehicles, and the surrounding driving environment. 

\subsection{Participants}
In the data collection phase, we recruited 22 participants: 9 people (8 male and 1 female) diagnosed with PD with an average age of 60 (PD 1-9) and 13 non-PD participants to participate in the experiment (NC 1-13); their demographic information is shown in Table \ref{tab:participant}. Because age can significantly impact one's motor abilities, we recruited participants of similar age in the experiment. We specifically included non-PD participants to provide a baseline of typical driving patterns. This baseline helped us compare with individuals with PD to identify both similar and irregular patterns, which informed the development of the subsequent clustering model.

The control group comprised retirees from a research institute and residents in the surrounding area, while people with PD were recruited from the neurology department of a hospital. The recruitment criteria for the participants included the following five requirements: 
1) has a motor vehicle driving license, 
2) has more than five years of driving experience, 
3) no motion sickness, 
4) no recent use of driving performance-related drugs,
5) The HY scale of people with PD does not exceed 3, while that of non-PD participants does not exceed 0.

People with PD were compensated in local currency equivalent to \$40 USD, which included payment for their participation in preliminary study, as mentioned in Section~\ref{interview participants}. Non-PD participants in this study were compensated in local currency, which is equivalent to \$15 USD. The recruitment and study procedure was approved by the local hospital's IRB. Each participant's engagement took approximately 50 to 70 minutes.

\subsection{Procedure}
Our experiment included preparatory work and three scenarios.
During the intermissions between each scenario, participants were provided enough rest until they reached optimal conditions. 
After each scenario, participants were asked to complete self-rating questions about their driving performance and the difficulty of the scenario settings. 

\subsubsection{Preparatory work} Upon arrival at the lab, the participants were asked to sign an informed consent form.
Next, we recorded the basic information and driving experience of the participants. We then set up a training scenario to help them become familiar with our system. The training scenario continued until participants indicated that they were ready for the formal driving tasks. Participants were asked to drive safely, obey traffic rules, and adhere to their typical driving habits in all scenarios.

\subsubsection{Driving scenarios}

\textbf{Scenario 1.} 
In this scenario, participants engaged in free driving in a city with a speed limit of 40km/h for approximately 10 minutes.  

\textbf{Scenario 2.} The scenario section was set up on a straight highway segment with a speed limit of 120km/h for approximately 10 minutes. This scenario was mainly a one-way, three-lane road. We implemented various emergency driving events to increase the difficulty and test the driving abilities of people with PD.

\textbf{Scenario 3.} The final scenario followed the same route as Scenario 1 but included emergency driving events. Participants were required to ensure safety by avoiding collisions with other vehicles and pedestrians crossing the road.

We employed a Latin Square design to minimize the impact of the experimental sequence on the results \cite{grant1948latin}.
Specifically, we used three experimental sequences: Scenario 1, Scenario 2, and Scenario 3; Scenario 2, Scenario 3, and Scenario 1; and Scenario 3, Scenario 1, and Scenario 2. Each group (PD and non-PD) followed its respective sequence in turn, ensuring balanced exposure to all experimental conditions within each group. All participants completed the three sequences except for P9, who began with Scenario 2 but had to withdraw from the experiment midway due to poor condition.

\subsubsection{Follow-up Interview} 
We further conducted a follow-up interview at the end of the data collection. Common questions included inquiries about the adaptability to the simulated driving equipment and differences between simulated and real-world driving. Non-PD participants were asked about the frequency of encountering these emergency driving situations in real life and the level of difficulty they experience when faced with them.
For people with PD, they were asked to report any irregular driving behaviors they had noticed during real-world driving.
\subsection{Data Processing Pipeline and Feature Extraction}

\begin{figure*}[!htbp]
  \centering
  \includegraphics[width=0.8\textwidth]{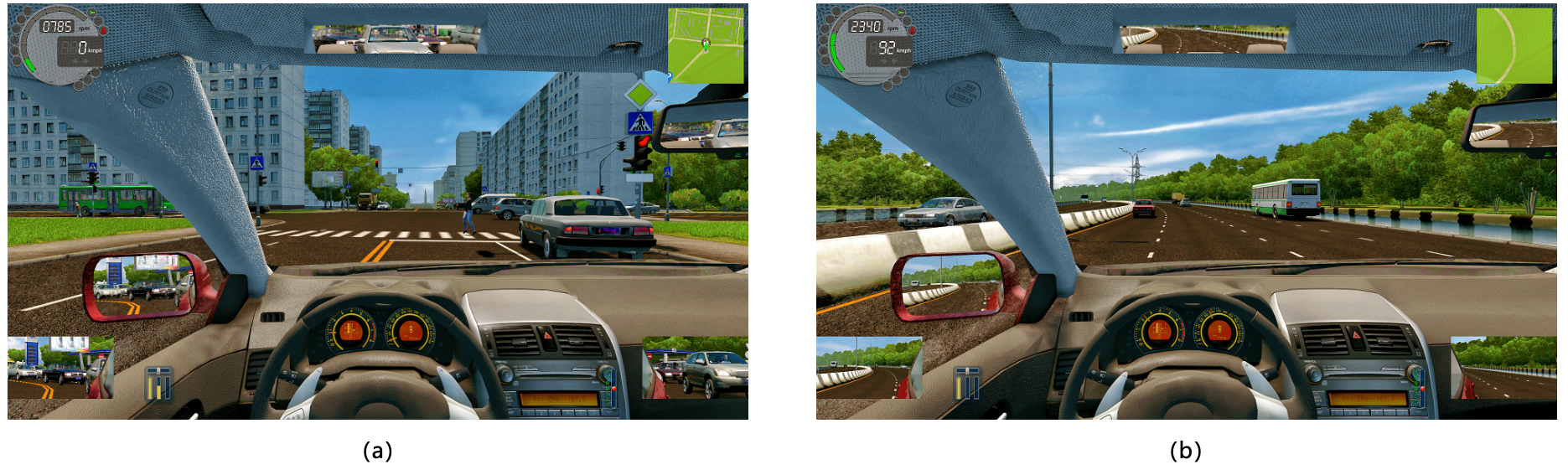}
  \caption{We set up two driving environments to collect real-time driving data from people with PD and non-PD participants: a city scene (shown on the left) and a highway scene (shown on the right).}
  \Description{The image consists of two panels, labeled (a) and (b), showing simulated driving environments as viewed from the driver's perspective.
  Panel (a): This image depicts a city driving scenario. The view includes tall buildings on both sides of the road, multiple vehicles on the road ahead, and visible traffic signs. The car's dashboard shows stationary status, as the speedometer indicates "0 km/h." The side mirrors and rearview mirror display surrounding traffic. A navigation map is visible in the top-right corner, indicating the route.
  Panel (b): This image represents a highway driving scenario. The road is surrounded by greenery, with fewer vehicles visible compared to the city environment. The car is in motion, as the speedometer shows "52 km/h." The side mirrors and rearview mirror display minimal traffic behind the vehicle. A navigation map is also present in the top-right corner, providing route guidance.}
  \label{fig:different_scenarios}
\end{figure*}

\subsubsection{Data collection and verification}
To collect the sensors (eye-tracking, steering wheel, and pedal sensor) data, we created a Python program on the computer. From our testing, we observed that eye movement data was occasionally inaccurate, with periods of reduced accuracy and decreased sampling rates. This issue was particularly pronounced in individuals with smaller eyes (PD1, PD9, NC7, and NC8). In real driving scenarios, head movements are more pronounced, and bright lighting conditions frequently occur, further complicating the collection of accurate eye movement data. To address this issue, we implemented a data verification step and introduced a 10-second buffer (as discussed in Section~\ref{driving alert}) to temporarily store data from each channel. Once the buffering was complete, we performed checks for timestamp, non-null values and verified the sampling rate to ensure data integrity. Across all 22 participants, the system tracking data experienced a loss of 30.8\% during the 700-minute experiment. So, we reduced the weighting of eye movement parameters in this study, utilizing fewer eye features to train the model.


\subsubsection{Data analysis}


To gain a deeper understanding of user behavior, the researcher reviewed the driving behavior video with the participants after they completed a stage of driving tasks. During this review, participants were asked to report their driving intentions, particularly during incidents such as accidents or traffic violations \cite{dawson2009ascertainment}. This section was divided into three main parts: a) Differences in data from each channel across various participants, b) Differences in behavioral performance, and c) Subjective intention and self-assessment of driving ability.

To summarize the driving ability of people with PD, particularly the stability of hand and foot control, we plotted the sensors' raw data for PD1, PD2, PD3, and NC1 during the straight-line driving stage (excluding crash incidents) in Figure \ref{fig:data}. Signal diagrams for all participants are provided in the Appendix. The x-axis in the figure represents time. The duration of the removed collision segments varies, leading to differences in the displayed time lengths. The y-axis represents the steering wheel angle, accelerator pedal depth, and brake pedal depth, with a consistent scale across all individuals.

\emph{\textbf{Differences in data from each channel across various participants}} 

\begin{figure*}
    \centering
    \includegraphics[width=0.8\linewidth]{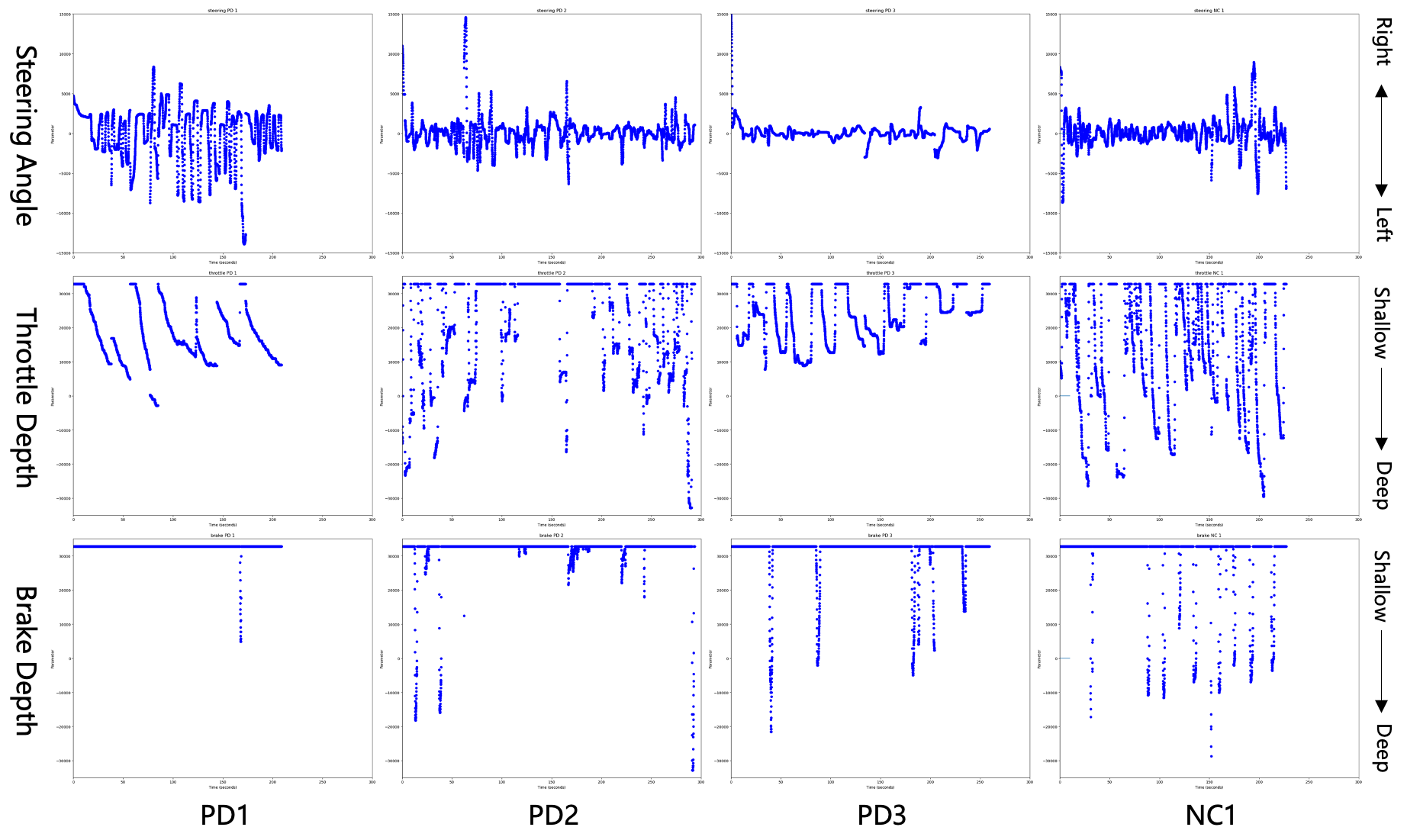}
    \caption{This chart compares the differences in the steering wheel, throttle, and brake data during straight-line driving among four participants: PD1, PD2, PD3, and NC1. (Data for all participants can be found in the Appendix.) In the chart, the horizontal axis represents time (in seconds), and the vertical axis represents steering wheel angle, throttle depth, and brake depth. The zero point on the vertical axis (center of the y-axis) indicates a steering wheel angle of 0, with positive values representing a right turn and negative values representing a left turn. For the throttle and brake, the highest values (at the top of the y-axis) represent fully released pedals, while lower values indicate deeper pedal engagement.}
    \Description{This figure contains a grid of 12 time-series plots, organized in 3 rows and 4 columns. The rows represent three driving metrics: Steering Angle, Throttle Depth, and Brake Depth, while the columns represent data from four participants: PD1, PD2, PD3 (all people with PD), and NC1 (a Non-PD participant). The x-axis of each plot represents time in seconds, and the y-axis represents the corresponding driving metric.
    Steering Angle (Row 1): The y-axis ranges from "Right" to "Left" steering angles. Plots for people with PD (PD1, PD2, PD3) show varying levels of fluctuation, indicating differing steering behaviors over time. The NC1 plot exhibits relatively stable steering with fewer abrupt changes than people with PD.
    Throttle Depth (Row 2): The y-axis ranges from "Shallow" to "Deep" throttle inputs. People with PD show a mix of shallow and deep throttle depths, with noticeable irregularities in throttle application. NC1 demonstrates more consistent throttle control, with smoother transitions over time.
    Brake Depth (Row 3): The y-axis ranges from "Shallow" to "Deep" brake inputs. Individuals with PD show intermittent brake application with varied depth, while NC1 shows a more controlled and consistent braking pattern. Each plot provides insight into the driving behaviors of people with PD versus the non-PD participants, highlighting variations in motor control and consistency during the driving task.}
    \label{fig:data}
\end{figure*}

\begin{itemize}
    \item \textbf{Steering data (excluding crash incidents)}: In our straight-line driving scenario, the lane is not perfectly straight. Consequently, we observed that non-PD individuals continuously adjust the steering wheel to fine-tune the car's direction. By comparing steering data from people with PD and non-PD individuals, we found that some people with PD can control the steering wheel quite accurately (e.g., PD2, Figure \ref{fig:data}) and avoid accidents (e.g., PD5, Figure \ref{fig:steering PD straight} in Appendix).
    However, other people with PD exhibit noticeable differences in their steering control. There are two main patterns: 1) PD1 demonstrates large steering wheel movements. In the interview, PD1 explained that he made conscious corrections when the car deviated from the lane center. However, due to PD-related impairments, he struggled to stop steering in time during turns, leading to over-correction in the opposite direction. This caused the vehicle to sway violently, increasing the times of a collision. 2) PD3 exhibited excessively rigid steering wheel operations. During data collection, he was transitioning from the "on" phase to the "off" phase of medication, showing severe muscle rigidity in both hands. Although PD3 did not crash as frequently as PD1, he struggled to make precise adjustments to the steering wheel, resulting in frequent boundary line violations during the simulation. In the subsequent interview, PD3 also reported that his current condition was poor.
    
    \item \textbf{Throttle data (excluding crash incidents)}: Through on-site observations and data analysis, we identified distinct patterns in how people with and without PD operated the throttle during straight-line driving—a task where maintaining a consistent speed is crucial to prevent accidents or being rear-ended. Two irregular throttle control patterns emerged. 1)In throttle modulation, NC1 frequently adjusted the throttle by alternating between pressing and releasing the pedal to maintain a stable speed, a behavior that PD2 could also replicate. In contrast, PD1 showed a very different pattern, holding the throttle in a steady, continuous press from shallow to deep over extended periods. This resulted in the car gradually accelerating until it eventually collided with an obstacle. In the follow-up interview, PD1 confirmed this behavior, attributing it to severe leg stiffness and slow movement, which were evident in his poor performance on the foot-tapping test, a part of the UPDRS-III assessment. This made it difficult for him to modulate the throttle frequently. 2)In throttle depth, PD3 never pressed the throttle beyond halfway throughout the session. He explained that his right leg was too slow and weak to apply sufficient force. Interestingly, non-PD participants NC2, NC7, and NC8 exhibited similar throttle patterns, although they attributed this to personal driving habits and differences in height. The accuracy and generalizability of this irregular pattern require further research.
    \item \textbf{Brake data (excluding crash incidents)}: We identified an irregular driving pattern in braking among people with PD. In straight-line driving scenarios, drivers often need to frequently apply the brakes to avoid sudden lane changes by other vehicles or multiple traffic incidents ahead. Data shows that most non-PD individuals frequently brake to keep their speed within a safe range and to avoid potential dangers. However, PD1 applied the brakes only once during nearly four minutes of straight-line driving. He rarely engaged in proactive deceleration, which contributed to the 20 collisions he experienced. In the subsequent interview, PD1 mentioned that he struggled to switch his right foot between the accelerator and brake pedals.
\end{itemize}

\begin{table*}[!htbp]
\centering
\begin{tabular}{cc}
\hline
\textbf{Features} & \textbf{Descriptions}\\
\hline
\textbf{Steering-based Features}\\
\hline
Steer Fluctuation Sum (Sum \& AbsSum) & The total sum or absolute value sum of steering fluctuation.\\
Fluctuation Times & The number of times the steering crosses zero. \\
Volume per Fluctuation & The total steering volume of one fluctuation. \\
Max Fluctuation & The maximum absolute value of steering changes. \\
Fluctuation Speed (Mean \& Max) & Average and peak speed of steering adjustment. \\
\hline
\textbf{Pedal-based Features}\\
\hline
Duration & The total time the throttle and brake were engaged. \\
Throttle-Brake Ratio & The ratio of throttle to brake duration. \\
Brake Times & The number of distinct braking events.\\
Throttle AUC & The sum of throttle values over time.\\
\hline
\textbf{Eye Movement-based Features}\\
\hline
Average Eye Speed & The average speed of the eye (X-axis, Y-axis, Trajectory). \\
Maximum Eye Speed & The maximum speed of the eye (X-axis, Y-axis, Trajectory). \\
Gaze Area Ratio & Ratio between gaze area and screen area.\\

\hline
\end{tabular}

\captionsetup{position=bottom} 
\caption{Features explored in this study} 
\Description{
    The table lists features related to steering, pedal usage, and eye movements, along with their descriptions:
    
    Steering-based Features:
    1. Steer Fluctuation Sum (Sum & AbsSum): The total sum or absolute value sum of the steering fluctuations.
    2. Fluctuation Times: The number of times the steering crosses zero.
    3. Volume per Fluctuation: The total steering volume in one fluctuation.
    4. Max Fluctuation: The maximum absolute change in steering.
    5. Fluctuation Speed (Mean & Max): The average and peak speed of steering adjustments.
    Pedal-based Features:
    1. Duration: The total time the throttle and brake were engaged.
    2. Throttle-Brake Ratio: The ratio of throttle time to brake time.
    3. Brake Times: The number of distinct braking events.
    4. Throttle AUC: The sum of throttle values over time.
    Eye Movement-based Features:
    1. Average Eye Speed: The average speed of the eye's movement (on the X-axis, Y-axis, and trajectory).
    2. Maximum Eye Speed: The maximum speed of eye movement (on the X-axis, Y-axis, and trajectory).
    3. Gaze Area Ratio: The ratio between the area of gaze and the screen area.
}
\label{tab:feature}
\end{table*}

\section{PANDA: Parkinson's Assistance and Notification Driving Aid}
Based on the driving data collected from the previous section and informed by conclusions drawn about effective alert strategies, we developed PANDA, a system designed to detect irregular driving behaviors in people with PD and provide early warning alerts. We use PANDA as a technology probe to further understand the experiences and design implications of real-time alert systems for individuals with PD.
\subsection{Drive Ability Detection}

\subsubsection{Feature extraction}
We used a sliding window method to process the multi-modal time-series data, with a window size of 10 seconds and an overlap of 50 percent.
This configuration was determined based on our interview results.
To balance between missing the optimal reminder time and having reminders too frequently, people with PD found that a 10-second window is ideal. While we fully respected their input, we used 10 seconds as a baseline and explored other parameter settings.
To accommodate the real-time requirements of the system, we initially tested four different sliding window lengths: 1, 3, 5, and 10 seconds, with each window overlapping by half of its length.

Based on the clustering results, we found that the 10-second sliding window provided the most accurate detection of irregular behaviors, making it the optimal choice for marking such events. 
Consequently, we selected a 10-second window for the final analysis. The features extracted from the eye tracker and pedal data provide a comprehensive assessment of the participant's visual-motion coordination and control strategies during tasks. These metrics can be used to evaluate attention, decision-making, reaction speed, and overall control efficiency, making them valuable for research in cognitive psychology, human factors, and driver behavior analysis.

\emph{\textbf{Eye Movement Features.}} The eye movement data provides insights into the participant's gaze behavior and eye movement dynamics. The key features extracted from this data include:

\begin{itemize}
    \item \textbf{Average Eye Speed (X-axis, Y-axis, Trajectory)}: The average speed of the eye. This reflects how fast the eye moves on the screen over time, which is important for analyzing attention shifts. Research has indicated that people with PD may show different eye-controlling abilities between horizontal and vertical movement. These differences may be an indicator of driving ability.
    \item \textbf{Maximum Eye Speed (X-axis, Y-axis, Trajectory)}: The maximum speed of the eye. This reflects the fastest eye movement on the screen, which is important for analyzing attention shifts, indicating the ability to catch the emergencies that occur on the road.
    \item \textbf{Gaze Area Ratio}: This is the ratio of the total area covered by the participant's gaze to the total screen area (1920 × 1080). This feature estimates how much of the visual space the participant engaged with and can reflect attentional spread or focus during the task.
\end{itemize}

\emph{\textbf{Steering Features.}} The steering data reflects the user's control over the vehicle's direction and their ability to maintain steady movements or adapt to changing circumstances. Key extracted features are:

\begin{figure*}
  \centering
  \resizebox{0.8\textwidth}{!}{\includegraphics{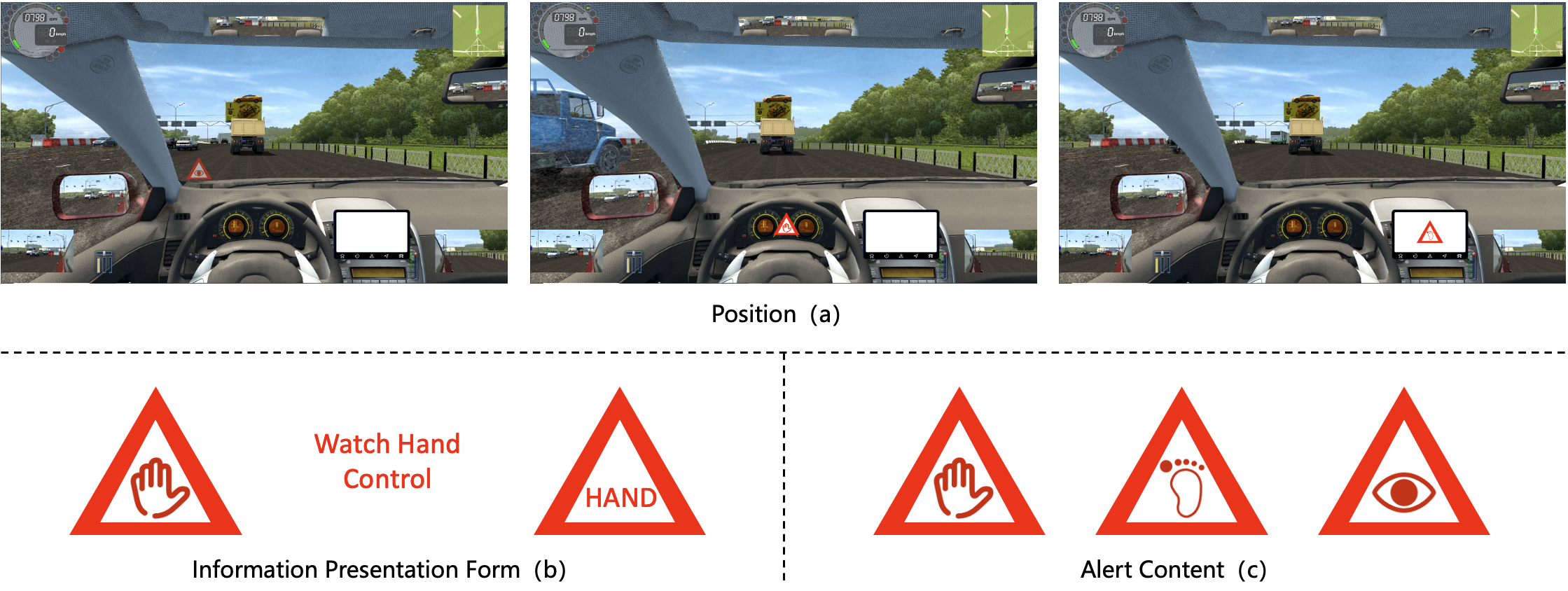}} 
  \caption{We investigated the effectiveness of visual alerts across three dimensions: position, information presentation form, and alert body part. The image illustrates our design setting: for the position, we evaluated the HUD, dashboard, and car center control screen from left to right in (a); for information presentation form, we used red warning triangles with graphic symbols, text only, and red warning triangles with text, from left to right in (b); and for alert content, we focused on hands, feet, and eyes, from left to right in (c).}
  \Description{
  This figure is divided into three sections labeled Position (a), Information Presentation Form (b), and Alert Content (c), each illustrating different aspects of an in-car alert system.
  Position (a): This section shows three images of the driving environment from the driver's perspective. Each image represents a different placement of an alert icon in the vehicle's interface. The left image shows the alert positioned on HUD. The center image shows the alert positioned in the middle of the dashboard. The right image shows the alert positioned on the car center control screen.
  Information Presentation Form (b): This section presents two variations of how the alert information is displayed. The first icon includes a red warning triangle with a hand symbol. The second icon only has the text "Watch Hand Control." The third icon is a simpler red warning triangle with the word "HAND" displayed inside.
  Alert Content (c): This section demonstrates three different types of alert content, all represented as red warning triangles but with different icons inside. The first icon includes a hand symbol, indicating a warning related to hand control. The second icon features a foot symbol, indicating a warning related to foot control. The third icon includes an eye symbol, suggesting a warning related to visual attention.
  }
  \label{fig:different places}
\end{figure*}

\begin{itemize}
    \item \textbf{Steer Fluctuation Sum (Sum and Absolute Sum)}: The total sum of steering fluctuation. This provides a relative fluctuation volume measuring steering activity, calculating the actual offset of the steering wheel from the beginning to the end of a slice. The absolute values of steering indicate the total fluctuation volume of a segment. This reflects the overall volume of steering corrections and adjustments, irrespective of direction.
    \item \textbf{Fluctuation Times}: The number of times the steering crosses zero (i.e., changes direction). This measures the variability of steering, which is important for identifying smooth versus erratic driving behavior.
    \item \textbf{Volume per Fluctuation}: The ratio of total steering volume to the number of fluctuation times. It represents the average amount of steering input per directional change, which can provide insights into steering efficiency.
    \item \textbf{Max Fluctuation}: The maximum absolute value of steering changes. This value identifies the largest steering adjustment made, which could indicate extreme maneuvers or corrections during the task.
    \item \textbf{Fluctuation Speed (Mean and Maximum)}: The average volume of steering adjustments over time. It provides a sense of how frequently and intensely the participant adjusts their steering. The peak fluctuation speed at which steering adjustments occur. This feature can indicate moments of rapid directional changes, reflecting reaction speed or compensatory behavior.
\end{itemize}

\emph{\textbf{Pedal Features.}} The pedal data (brake and throttle) provides insights into the participant's acceleration and deceleration behavior, which is critical for understanding their control strategy and response to the environment. Extracted features include:

\begin{itemize}
    \item \textbf{Duration (Throttle and Brake)}: The total time the throttle and brake were engaged, representing the duration of acceleration and deceleration. A longer throttle duration indicates prolonged acceleration phases.

    \item \textbf{Throttle-Brake Ratio}: The ratio of throttle to brake duration. This feature is useful for understanding the balance between acceleration and deceleration. A high ratio suggests aggressive driving, while a low ratio could indicate cautious behavior.

    \item \textbf{Brake Times}: The number of distinct braking events. This is calculated by counting the times when the participant lifts off the brake completely between braking events, reflecting braking frequency.

    \item \textbf{Throttle Area Under the Curve (Throttle AUC):} The sum of throttle values over time, indicating the total amount of throttle applied during the session. Since the behavior of engaging the throttle is not as distinct as braking, with clear phases of pushing and releasing, it involves continuous engagement. Therefore, we use the throttle AUC (Area Under the Curve) to indicate changes in throttle behavior. This value has a strong connection with the speed and acceleration. A higher AUC reflects more intensive acceleration behavior.
\end{itemize}

\subsubsection{Clustering of irregularities}

\emph{\textbf{Data Preprocessing.}}
Preprocessing steps were undertaken to ensure data quality and consistency in preparing the driving-related time-series data for clustering analysis. 
Normalization of feature values was performed using min-max scaling, which transformed all features to a common scale between 0 and 1. 
This step was critical for ensuring that no feature with a larger range disproportionately influences the clustering results, especially in distance-based clustering algorithms like KMeans.

\emph{\textbf{Clustering with KMeans.}}
KMeans clustering was selected as the primary method for grouping driving behaviors based on feature similarities. 
Specifically, the dataset was divided into two clusters, hypothesized to represent regular and irregular driving patterns. 
The number of clusters (n=2) was chosen based on the need to distinguish between these two behavioral groups. The KMeans algorithm works iteratively by assigning data points to clusters based on proximity to centroids, which are recalculated to minimize intra-cluster variance. 
After clustering, each sample was labeled with its corresponding cluster assignment, enabling further analysis of the driving patterns associated with each group.

\emph{\textbf{Model Persistence and Prediction.}}
To ensure that the clustering model could be reused for future analyses without retraining, the KMeans model was saved using a serialization method. 
This allowed the trained model to be stored and later reloaded for real-time application or batch prediction of new driving data. 
This approach facilitates the dynamic use of the model in real-world scenarios where continuous assessment of driving behavior may be required. 
The stored model was later utilized to predict the cluster assignment of specific test samples, confirming the model's reliability in detecting irregular driving patterns.

\begin{table*}[t]
\centering

\begin{tabular}{c|ccccccccc}
\hline
\ & \textbf{P2}  & \textbf{P4} &\textbf{P5} &\textbf{P6}&\textbf{P7}&\textbf{Average}&\textbf{STD}\\
\hline
\multicolumn{8}{c}{How much do you like this visual cue? Please rate from 1 (dislike very much) to 5 (like very much).}\\
\hline
HUD & 3 & 4 & 4 & 3 & 5 & \textbf{3.8} & 0.8\\
Dashboard & 4 & 3 & 2 & 1 & 4 & 2.8 & 1.2\\
Center Control Screen & 5 & 2 & 3 & 1 & 3 & 2.8 & 1.3\\
\hline
\multicolumn{8}{c}{To what extent do these visual cues affect driving? Please rate from 1 (not affected at all) to 5 (very affected)}\\

\hline
HUD  & 1 & 2 & 1 & 3 & 2 & \textbf{1.8} & 0.7\\
Dashboard & 1 & 3 & 3 & 4 & 1 &2.4 &1.2\\
Center Control Screen & 1 & 4 & 2 & 5 & 4 &3.2 &1.5\\

\hline
\end{tabular}

\captionsetup{position=bottom} 
\caption{Result for alert preference study Scenario 1 on 5-point Likert scale. More people preferred having visual alerts displayed on the HUD, and most believed that presenting visual alerts on the HUD would minimally impact normal driving.} 
\Description{
    The table shows ratings from five participants (P2, P4, P5, P6, and P7) regarding two questions.
    Question 1: How much do you like this visual cue? Please rate from 1 (dislike very much) to 5 (like very much).
    HUD (Heads-Up Display):
    Participants' ratings: 3, 4, 4, 3, 5
    Average: 3.8, Standard Deviation (STD): 0.8
    Dashboard:
    Participants' ratings: 4, 3, 2, 1, 4
    Average: 2.8, STD: 1.2
    Center Control Screen:
    Participants' ratings: 5, 2, 3, 1, 3
    Average: 2.8, STD: 1.3
    Question 2: To what extent do these visual cues affect driving? Please rate from 1 (not affected at all) to 5 (very affected)
    HUD:
    Participants' ratings: 1, 2, 1, 3, 2
    Average: 1.8, STD: 0.7
    Dashboard:
    Participants' ratings: 1, 3, 3, 4, 1
    Average: 2.4, STD: 1.2
    Center Control Screen:
    Participants' ratings: 1, 4, 2, 5, 4
    Average: 3.2, STD: 1.5
    }
\label{tab:visual_result}
\end{table*}

\subsection{Alert Strategy Design}
\label{alert design space}

In the preliminary study (Section~\ref{driving alert}), we explored the design space of alert methods for a real-time driving reminder system in collaboration with individuals with PD. Our primary focus was on how to design alerts through three channels: visual, auditory, and tactile for people with PD. We examine which design elements need to be considered for these channels and how to differentiate the design of alerts for various driving scenarios. 

In our \textbf{visual alert} design, we considered three key dimensions: position, information presentation form, and alert content. For position, we designed alerts placed in the head-up display (HUD) \cite{langner2016traffic}, on the dashboard \cite{barber1994psychological}, and on the car center control screen \cite{lin2023evaluating}. For the information presentation form, we designed alerts with text only, red warning triangles with text, and red warning triangles with graphic symbols \cite{winkler2018warn}. For alert content, we included alerts for hands, feet, and eyes. Our visual alert design is shown in Figure \ref{fig:different places}.

As for \textbf{audio alert} design, we included three types of audio alerts under various common on-road situations in both healthy and elderly drivers with diseases: a warning sound only, a warning sound with what to do and a warning sound with both what to do and why it should be done \cite{koo2015did}. The on-road situations included starting, encountering traffic signals, making turns, lane keeping, overtaking, speed control, backing up, and curving, according to this paper \cite{dawson2009ascertainment}.

Regarding \textbf{tactile alert}, nearly all participants ruled out the use of the tactile channel. As P4 mentioned: "\textit{The tactile channel cannot convey as much detailed information as the visual and auditory channels. When I receive a tactile alert, I find it difficult to discern what action is required of me.}"

In the co-design process with people with PD, we identified two key driving scenarios that necessitate alerts. The first scenario involves detecting irregular driving patterns in people with PD; timely reminders in these instances can help them adjust their physical state or decide to make an emergency stop, thus preventing accidents. The second scenario pertains to situations requiring complex information processing or those with a higher risk of traffic accidents, where scenario-specific alerts are crucial to help the driver focus on critical issues.

Building on the insights from the preliminary interviews and the alert design space, we developed three modes in PANDA for the subsequent user study: the visual alert test mode, the audio alert test mode, and the system experience mode. The visual alert test mode randomly presents nine types of visual alerts (three positions × three alert contents) every 30 seconds, as described in Section~\ref{alert design space}. The audio alert test mode plays three types of audio alerts for common on-road situations, also detailed in Section~\ref{alert design space}. The system experience mode automatically detects irregular driving patterns and presents alerts during free driving. To protect users' privacy, we designed a feature that allows them to turn alert notifications on or off at any time.


\section{User Study}
After building PANDA, we designed this experiment to validate its usability through user study with 5 people with PD. We use PANDA as a technology probe to explore the alert preferences and driving experiences of individuals with PD, as well as the design implications for a real-time driving alert system.

\subsection{Participants}
We recruited 5 people with PD (P2, P4, P5, P6, and P7 in Table~\ref{tab:patient_info}) took part in this user study. 
Participants were compensated in local currency equivalent to \$40 USD in user study. The recruitment and study procedure was approved by the local hospital's IRB. Each study took about 70 minutes.



\subsection{Study Procedure}
The user study procedure contains three phases: 1) Real-time experience with PANDA, 2) Alert method preference, and 3) Semi-structured interview.

\subsubsection{Phase 1: Real-time experience with PANDA}[30 Minutes]
Participants drove with PANDA activated. After the experience, they were asked the following questions: How satisfied were they with the system? After receiving an alert, could they recognize their issue and subsequently adjust to a better driving state? Which three driving situations did they prefer to use this system in? Which three driving situations would they avoid using this system?


\subsubsection{Phase 2: Alert method preference}
\textbf{Scenario 1} [10 Minutes]
Drivers navigating city roads encountered nine types of visual alerts (three positions × three alert contents), which appeared randomly every 30 seconds. Afterward, their preferences regarding the position of the visual alerts and the distracting degree of these alerts were measured using a 5-point Likert scale \cite{joshi2015likert}. Several questions were posed, including whether they felt different alert contents were necessary and why, as well as their preferred format for presenting alert information.

\textbf{Scenario 2} [10 Minutes]
Participants drove on city roads and, while performing each on-road situation, encountered three different types of audio alerts in a Latin Square sequence. Afterward, they were asked to rank their preference among the three types of audio alerts.

\subsubsection{Phase 3: Semi-structured interview}[20 Minutes]
The interview explored two main questions: First, are driving statistical parameters related to hand, foot, and eye movements useful for people with PD and how should they be presented? Second, in which scenarios would people with PD prefer to use our system, such as driving alone, driving with others, highway driving, or city driving?

\begin{figure*}[!htbp]
    \centering
    \includegraphics[width=0.8\linewidth]{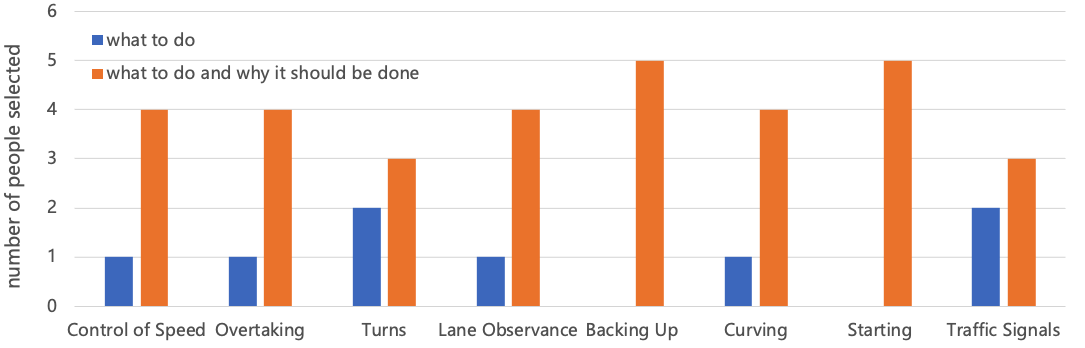}
    \caption{In alert preference study Scenario 2, we evaluated the preferences of people with PD for audio alerts during different driving tasks. Upon entering each driving scenario, participants were asked to choose between short, simple audio alerts ("what to do") and longer, more detailed audio alerts ("what to do and why it should be done"). The results showed that people with PD consistently preferred detailed audio alerts ("what to do and why it should be done") across all driving tasks.}
    \Description{The chart compares two types of information preferences for various driving tasks, as selected by participants: "what to do" (represented by blue bars) and "what to do and why it should be done" (represented by orange bars). The y-axis represents the number of people who selected each type of information, while the x-axis lists specific driving tasks.
    Control of Speed: 1 person preferred "what to do." 4 people preferred "what to do and why it should be done."
    Overtaking: 1 person preferred "what to do." 4 people preferred "what to do and why it should be done."
    Turns: 2 people preferred "what to do." 3 people preferred "what to do and why it should be done."
    Lane Observance: 1 person preferred "what to do." 4 people preferred "what to do and why it should be done."
    Backing Up: 0 person preferred "what to do." 5 people preferred "what to do and why it should be done."
    Curving: 1 person preferred "what to do." 4 people preferred "what to do and why it should be done."
    Starting: 0 person preferred "what to do." 5 people preferred "what to do and why it should be done."
    Traffic Signals: 2 people preferred "what to do." 3 people preferred "what to do and why it should be done."}
    \label{fig:audio preference}
\end{figure*}

\subsection{Data Analysis}
All information from the user study was meticulously documented, and the entire study was fully recorded in audio. The semi-structured interviews in the user study were analyzed using thematic analysis \cite{braun2006using}. 
Firstly, all audio recordings were transcribed into text and cross-checked against handwritten notes to serve as the data input. Secondly, two researchers independently employed thematic analysis to identify, analyze, and interpret themes in the data. Any conflicts in theme identification were resolved through discussion and consensus. The analysis focused on two key themes: feedback of the system and acceptance of scenarios. The Cohen's Kappa coefficient for inter-rater reliability between the two researchers was 0.667. Further details on the thematic analysis process are provided in Table~\ref{tab:thematic analysis user study} in the Appendix.

\subsection{Findings}

\subsubsection{Insights for overall experience with real-time driving alert system}

\textbf{A real-time driving alert system is helpful for people with PD. When they receive an alert, they can quickly adjust their driving behavior, reducing the risk of potential dangers.} During the real-time alert phase, every person with PD received at least one real-time warning. More than half of these alerts were deemed effective by the patients, who reported issues such as difficulty maintaining speed or unstable steering. All participants found the system highly useful. Since people with PD experience more pronounced fluctuations in their physical condition compared to non-PD individuals, the system helps ensure their safety. P6 noted, "\textit{people with PD have noticeable on-off period effects. During the 'on' period, I can operate the car well, but in the 'off' period, my body becomes stiff, and movement is difficult. Since it's impossible to predict when the switch will occur, this system helps me assess my physical state and take necessary actions in advance.}" P7 mentioned, "\textit{Because I need to take medication frequently, the system's alerts are helpful when I start to feel drowsy.}"

\subsubsection{The privacy protection feature is especially beneficial for people with PD. People with PD prefer to use the real-time driving monitoring system when driving alone rather than when they have passengers in the car.}
When driving alone, there is no co-driver to assist with observations and reminders, which can increase the risk of drowsiness and distraction.
Conversely, having others in the car can make triggering reminders more problematic. It may reveal personal information and reduce self-confidence. P5 shared, “\textit{If the system issues an alert while my boss, colleagues, or friends are in the car, everyone will know about my condition. It's embarrassing, and they might even tell others about the warnings, which would be humiliating.}”
P4 noted, “\textit{My family members often give me driving instructions when they're in the car. Frequent reminders could provide them with more reasons to criticize me, leading to potential conflicts.}”

\subsubsection{Alert method preference}

\textbf{Insight 1: Visual alerts should be placed closer to the driver's line of sight, primarily using graphical information. It is helpful to provide individuals with PD precise information about which specific body parts require adjustment.} 

As is shown in Table \ref{tab:visual_result}, more participants preferred having visual alerts displayed on the HUD, with an average rating of 3.8. Additionally, most people believed that presenting visual alerts on the HUD would minimally impact their normal driving, giving it an average score of 1.8. The placement of visual prompts should be closer to the driver's line of sight. For individuals with PD, head and eye movement capabilities are generally weaker, and information processing speed is slower compared to non-PD individuals. Positioning visual prompts closer to the driving line of sight ensures that people with PD do not become distracted by having to look away from the road, which could lead to more severe driving issues. P4 mentioned: "\textit{When looking at the dashboard alerts, I need to lower my head and shift my gaze from directly ahead to the dashboard. When checking alerts on the car's central display, I have to adjust my gaze to the right and downward.}" This opinion aligned with survey results, more people believed that HUD was a better method of providing prompts and results in the least amount of distraction while driving.

Incorporating information about which body part (hand, foot, or eyes) needs more attention in the alerts was highly appreciated by all people with PD. They generally felt that identifying the specific body part requiring improvement helped them better direct their physical adjustments. Regarding the format of the information presentation, all participants preferred red warning triangles with graphic symbols over red warning triangles with text, with text only being the least effective. P6 remarked, "\textit{Understanding text takes more time than interpreting icons, and reading text while driving can disrupt normal driving. This system should include training for drivers to become familiar with these icons before implementation, and during driving, different shapes and positions of icons can be used to differentiate various alert messages.}"


\textbf{Insight 2: When designing the audio alert content, factors such as driving speed, frequency of occurrence, and the level of risk in each driving scenario should be considered to provide precise and timely instructions.} When discussing the design of voice prompts, P5 suggested, "\textit{The prompts should be concise and clear. In situations where alerts are triggered frequently, long voice prompts would become annoying. Moreover, during high-speed driving or in hazardous conditions, lengthy and detailed prompts could cause me to miss the optimal time for corrective action.}" P6 noted, "\textit{Voice prompts should be integrated with navigation systems and in-car cameras to assess the external environment, making the alerts more accurate. If detailed voice prompts fail to clearly identify the danger or provide specific instructions, they may mislead the driver. In such cases, a general warning would be more effective.}". All participants agreed that a simple alert sound is not sufficient. P6 remarked, "\textit{An alarm without specific instructions leaves me unsure of what action to take.}"


\begin{figure*}[!htbp]
    \centering
    \includegraphics[width=0.8\linewidth]{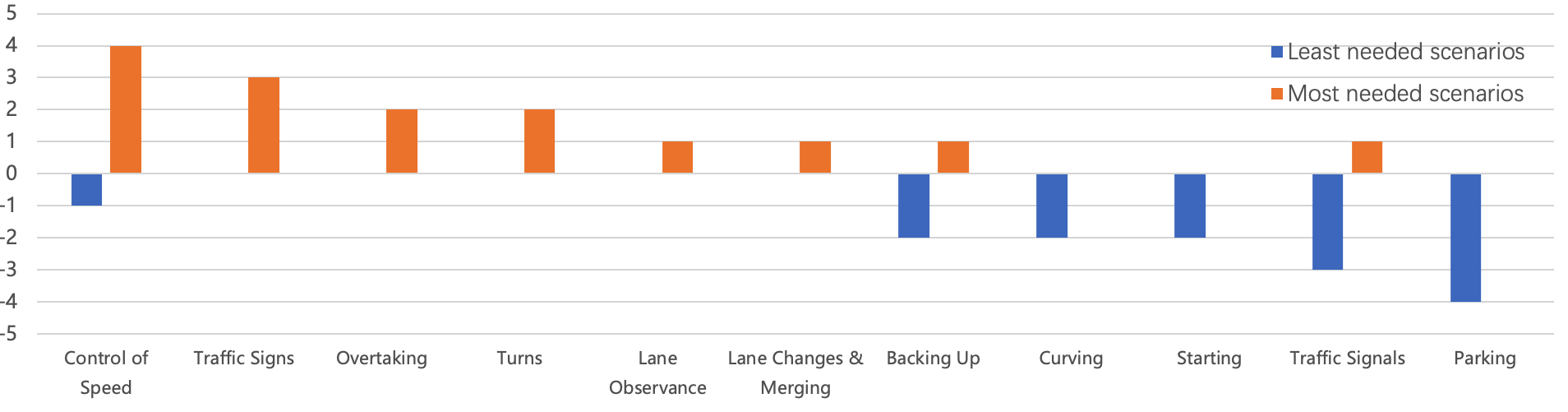}
    \caption{Each participant chose three most needed and three least needed scenarios for our alert system in the interview: The scenarios where people with PD most need are speed control and encountering traffic signs. The scenarios where they need it the least are parking and encountering traffic signals.}
    \Description{The chart illustrates the scenarios where audio alerts are considered either most needed or least needed by participants. The y-axis represents the number of participants, with positive values indicating "most needed scenarios" (orange bars) and negative values indicating "least needed scenarios" (blue bars). The x-axis lists specific driving scenarios.
    Control of Speed: 4 participants selected it as a "most needed scenario." 1 participant selected it as a "least needed scenario."
    Traffic Signs: 3 participants selected it as a "most needed scenario." No participants selected it as a "least needed scenario."
    Overtaking: 2 participants selected it as a "most needed scenario." No participants selected it as a "least needed scenario."
    Turns: 2 participants selected it as a "most needed scenario." No participants selected it as a "least needed scenario."
    Lane Observance: 1 participant selected it as a "most needed scenario." No participants selected it as a "least needed scenario."
    Lane Changes and Merging: 1 participant selected it as a "most needed scenario." No participants selected it as a "least needed scenario."
    Backing Up: 1 participant selected it as a "most needed scenario." 2 participants selected it as a "least needed scenario."
    Curving: No participants selected it as a "most needed scenario." 2 participants selected it as a "least needed scenario."
    Starting: No participants selected it as a "most needed scenario." 2 participants selected it as a "least needed scenario."
    Traffic Signals: 1 participant selected it as a "most needed scenario." 3 participants selected it as a "least needed scenario." 
    Parking: No participants selected it as a "most needed scenario." 4 participants selected it as a "least needed scenario."}
    \label{fig:need_scenarios}
\end{figure*}

\subsubsection{The scenarios where people with PD most need driving alerts for irregularities}

During the preliminary study, interviews revealed that people with PD require clear, accurate, and context-specific alerts tailored to different driving scenarios. Previous research has shown that as PD progresses, patients develop consistent patterns of motor decline. These fixed motor patterns can lead to poor performance in certain scenarios that require specific motor functions, while their performance may be acceptable in other scenarios. Therefore, we believe that a driving alert system for individuals with PD should accurately distinguish between critical and less critical scenarios. In the most critical scenarios, the system should increase the frequency and intensity of alerts, while reducing them in less critical scenarios.

To identify these scenarios and understand the reasons behind them, we included a feasibility validation phase where users drove through eleven different scenarios mentioned in a previous work \cite{dawson2009ascertainment} in a random sequence, while triggering system alerts. In subsequent semi-structured interviews, participants were asked to select the three scenarios where they felt real-time alerts were most and least necessary. The result is shown in Figure \ref{fig:need_scenarios}. The overall results showed that Control of Speed and Traffic Signs are the top 2 most needed, while Traffic Signals and Parking are the least. The following explains why patients made these choices:

\textbf{Control of Speed.}
People with PD identified this scenario as the one most in need of real-time alerts. Our system's feature extraction showed that some people with PD have poorer throttle and brake control compared to the non-PD group. P5, who experiences poor coordination in the right lower limb, explained: “\textit{My right leg often becomes very stiff, and my foot movements slow down and lose strength. Since developing the disease, I can no longer press the pedal as deeply or accurately as I used to. I often realize only halfway through pressing the pedal that I haven't pressed it as deeply as I intended, and then I need to press it again. Additionally, my speed in switching my foot between the accelerator and brake has decreased.}” With impaired control over the accelerator and brake, people with PD find it challenging to monitor their speed on the dashboard, making it difficult to maintain appropriate speeds. Meanwhile, driving too fast or too slow can lead to severe traffic accidents.

\textbf{Traffic Signs.}
Encountering traffic signs was the second most frequently chosen scenario by people with PD. These traffic signs include stop signs, no right turn on red signs, one-way street signs, pedestrian crossing signs, and no-entry signs. Previous studies have shown that people with PD tend to have a reduced Useful Field of View (UFOV) \cite{classen2009useful}, and in experiments, they identified fewer landmarks and traffic signs compared to control groups \cite{uc2006impaired}. When asked why they made this choice, P5 explained: "\textit{I feel that while concentrating on driving, I can't pay attention to multiple factors on the road as I could before I got sick. Missing these traffic signs can have serious consequences, often leading to major traffic violations.}"

\textbf{Overtaking, Turns, Lane Changes \& Merging and Curving.}
people with PD often experience resting tremors or rigidity, and when these symptoms affect the hands, they can significantly impair steering control. Participants ranked overtaking and turning as the third most critical scenarios, and lane changing \& merging as the fourth, for needing a driving alert system. In these situations, people with PD need to change lanes or turn, requiring both hands to maintain precise control of the steering wheel. P6, who experienced significant tremors in the left hand, shared: "\textit{I usually steer with just my right hand. When making sharp turns, it's difficult to manage without using my left hand. However, when I use both hands, my control over the wheel's stability is noticeably worse than with just one hand. Additionally, during lane changes and turns, I often need to consider vehicles and pedestrians outside my field of vision, which presents a challenge for me.}" 
However, driving on curved roads, which also requires steering, was voted as the third least necessary scenario for alerts. When asked why, P5 explained: "\textit{On curved roads, I only need to focus on staying within the lane and don't have to pay as much attention to the behavior of other vehicles. It's simpler.}"

\textbf{Lane Observance.}
Keeping the vehicle within the lane was relatively less in need of additional alerts for people with PD. P4 mentioned: "\textit{Nowadays, many cars are equipped with lane-keeping features, which are already sufficient to help me stay in the lane.}"

\textbf{Starting.}
The start-up phase was ranked as the third least necessary scenario for alerts by people with PD. P2 shared: "\textit{Before I start driving, I always check my surroundings, and this habit ensures my safety during the start-up phase. Additionally, since the vehicle is moving slowly at this stage, it's easier to control.}"

\textbf{Traffic Signals.}
Encountering traffic lights was ranked as the second least necessary scenario for alerts by people with PD. P4 explained: "\textit{Aside from turning, intersections with traffic lights typically involve low-speed, straight-line driving. Plus, there's plenty of time to observe the surroundings while waiting for the green light.}"

\textbf{Parking and Backing Up.}
Parking and backing up were ranked as the least and third least necessary scenarios for alerts by people with PD, for similar reasons. P7 explained: "\textit{When parking or backing up, the vehicle is moving at a low speed, making serious accidents less likely. More importantly, my car is equipped with parking sensors and a reverse-assist system, which are fully capable of helping me with these maneuvers.}"

\section{DISCUSSION}
In the discussion, we further discuss future opportunities regarding system optimization directions, comprehension of system alerts by individuals with PD, and transferring our work to real-world driving environments.

\subsection{System Optimization Directions}
From the research, we found that there is significant variability among people with PD, so the system should offer personalized customization features. These differences manifest in two key aspects: 1) the progression of the disease, typically described by the Hoehn-Yahr scale~\cite{modestino2018hoehn}, and 2) the dominant symptoms, which can vary even among people at the same Hoehn-Yahr stage~\cite{martinez2013expanded}. Some people may experience bradykinesia (slowness of movement), while others may exhibit resting tremors, which often appear on one side of the body in the early stages~\cite{Willis2022, Armstrong2020}. Given that driving (specifically in left-hand drive cars) places the highest demand on the right leg, and that one stable arm is usually sufficient to control the steering wheel, both the progression of the disease and the dominant symptoms can have a significant impact on driving ability~\cite{classen2014driving, crizzle2013postural}. Therefore, a personalized design approach is essential.

Furthermore, since eye-tracking data collection methods may depend on factors such as eye size or the use of corrective lenses~\cite{brunye2019review, wolf2023eye,bigham2021fly,fan2020eyelid,tian2024designing}, determining the driving condition of people with PD based solely on eye-tracking data may not be reliable for everyone. As a result, the system should offer two modes: one that incorporates eye-tracking data and one that does not, in order to enhance robustness. In addition, during the collaborative design process with people with PD, we explored tactile feedback methods. However, compared to visual and auditory cues, tactile feedback lacked the richness needed to differentiate between various types of irregular behavior, and it was challenging to calibrate the intensity to capture attention without interfering with driving. As a result, this method was not presented in user experiments. The use of tactile feedback for alerting people with PD to driving irregularities remains an area requiring further research.

Finally, our system has the potential to be adapted for broader use, assisting individuals whose driving abilities are impaired by similar symptoms. Since our system is capable of detecting irregular driving patterns caused by PD—such as stiffness and tremors in the limbs, reduced attention span, and slower reaction times—it could also be applied to older adults \cite{li2019fmt} and individuals affected by other conditions with similar manifestations, such as essential tremor, Alzheimer's disease, and stroke \cite{li2021choose,pei2023embodied,han2024co,luo2024exploring1}.

\subsection{Comprehension of System Alerts by Individuals with PD}
During the experiment, when people with PD received system alerts, we asked them what they believed had triggered the alert. Since most participants didn't have a computer science background and we didn't explain the underlying logic of the system's alert algorithm, their interpretations of the reasons behind the alerts sparked our thinking. Sometimes, their reasoning matched our design criteria (e.g., not braking for a certain period or excessive steering wheel movement), but on other occasions, they attributed the alerts to factors outside the system's scope (e.g., a traffic accident ahead).

This gap in comprehension highlights a critical aspect of human-centered AI in assistive driving systems. Human-centered AI emphasizes the importance of creating an AI system that is reliable, safe, and trustworthy \cite{shneiderman2020human}. To improve the quality of life for people with PD, it also underscores the necessity for AI systems to address real-life human challenges, such as eating and walking \cite{al2022smart, bachlin2010wearable, canning2020virtual}. In assistive driving systems, this approach is particularly vital, as drivers must trust and act on the system's alerts to make real-time decisions. If alerts are not communicated clearly, the system’s effectiveness can be compromised, potentially leading to incorrect decisions and new safety risks, as shown by the participants’ misunderstandings.

To address this, we employed a data-driven ML pipeline. This pipeline extracts features from specific dangerous behaviors and trains the model based on these features. By doing so, the system is able to link alerts to the corresponding behaviors, ensuring that the reasons behind each alert are more directly tied to observable, understandable driving patterns. This approach enhances the clarity of alerts and aligns the system's actions with users' expectations, ultimately improving the system's effectiveness and user trust.

\subsection{Transferring This Work to Real-World Driving Environments}
Firstly, the evaluation parameters used in our system can be obtained from existing equipment in real-world driving environments, eliminating the need for additional hardware. Eye movement data can be captured using in-vehicle cameras designed for detecting driver fatigue and distraction through computer vision techniques~\cite{wang2021survey, mavsanovic2019driver}, while pedal and steering wheel data can be retrieved from the car's CAN bus \cite{lokman2019intrusion}. Thus, deploying the system's algorithm on the vehicle's onboard computer will involve relatively low future costs in terms of time and money. Secondly, in practical use, the system can be integrated with existing vehicle fatigue and distraction monitoring systems~\cite{koch2023leveraging, sikander2018driver}, which helps mitigate some of the interference from irregular driver states on the system's algorithms.

Our system has proven the feasibility of using algorithms to monitor the driving behavior of people with PD. However, transitioning to a real-world driving environment presents challenges due to differences between driving simulators and actual driving conditions (e.g., real vehicles have different steering wheel and pedal resistance parameters, real-world driving offers a wider visual field, and the driving environment can vary with road conditions and bumps). To successfully implement the model in real-world settings, we must collect larger-scale driving data from people with PD in controlled on-road environments, alongside data from non-PD drivers, to refine and validate the computational parameters. The system's effectiveness should then be verified by people with PD who have been assessed as fit or unfit to drive by official driving evaluation agencies before it is applied to real-world driving scenarios.

\section{Limitation}

In our research, we designed and evaluated PANDA, a system aimed at helping individuals with PD drive safely and confidently. However, the relatively small participant size may limit the generalizability of our findings to the broader PD population. Nevertheless, we anticipate that additional data collection will enhance the robustness of our alert model.
For safety and ethical reasons, we utilized a simulated driving approach to investigate the driving patterns of individuals with PD. Incorporating real-world driving scenarios has the potential to further advance our work toward practical applications in everyday driving settings.
Furthermore, our invitation-based recruitment strategy may have inadvertently excluded participants with poorer driving abilities, potentially limiting the comprehensiveness of our dataset.
During the driving data collection phase, we aimed to implement a fully balanced Latin Square design to ensure comprehensive data coverage. While the premature withdrawal of a participant with PD and the non-PD participant count not being a multiple of the conditions resulted in a slight imbalance, we believe the design still provides meaningful insights and maintains its capacity to support our analysis effectively. 
Despite the limitations mentioned above, we believe our work made a novel contribution in exploring the challenges faced by people with PD in driving and have successfully investigated the feasibility of analyzing their driving status in real-time under the given conditions.

\section{CONCLUSION}
Many emerging technologies are dedicated to improving the quality of life for people with PD. In this work, we conducted a series of investigations aimed at helping people with PD drive more confidently and safely. In the preliminary study, we thoroughly identified existing practices, perceptions, and challenges of people with PD regarding driving to guide design decisions in the development of PANDA. 
Using data collected from a driving simulator experiment, we trained a machine learning model to detect driving irregularities driving behaviors in people with PD and developed PANDA. Finally, user experiments provided valuable guidance on the design of alert methods and validated the feasibility of the real-time alert system. Our work represents a first step toward the real-time, quantitative analysis of driving behavior in people with PD.

\begin{acks}
This work was supported by the National Science and Technology Major Project of China (Grant No. 2022ZD0118002), the National Natural Science Foundation of China (Grant Nos. 62332017 and 62202456), and the CAS Project for Young Scientists in Basic Research (Grant No. YSBR-040). We would like to express our special gratitude to all the participants in our study.
\end{acks}
\bibliographystyle{ACM-Reference-Format}
\bibliography{reference/sample-base}


\clearpage
\onecolumn
\appendix
\section{Data of All Participants}
Due to the limited space in the main text, we present the driving performance data collected from all participants here, as described in the data collection procedure (Section~\ref{data collection}), to supplement the completeness of the data.
\subsection{Steering Control Performance of People with PD}
\begin{figure}[H]
    \centering
    \includegraphics[width=0.7\linewidth]{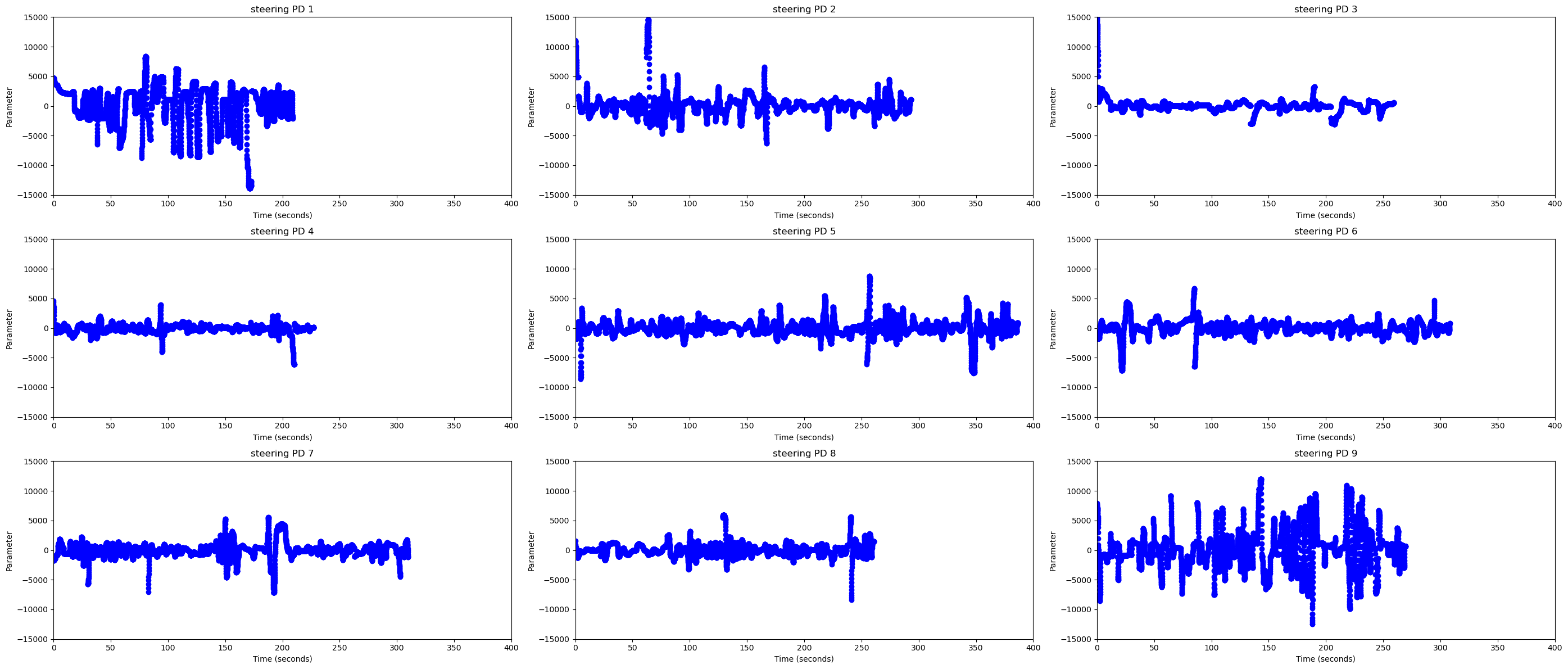}
    \caption{This figure sequentially presents the steering control performance of PD 1 to PD 9 during the data collection process. In each chart, the horizontal axis represents time in seconds, while the vertical axis represents the steering wheel angle. The zero point on the vertical axis (center of the y-axis) indicates a steering wheel angle of 0, with positive values representing a right turn and negative values representing a left turn.}
    \Description{This figure consists of nine line plots arranged in a 3 × 3 grid. Each plot represents the steering behavior of a person with PD over time, labeled PD 1 through PD 9. The x-axis in all plots represents time in seconds (ranging from 0 to 400), and the y-axis represents a parameter related to the steering angle, with values ranging from -15,000 to 15,000.}
    \label{fig:steering PD straight}
\end{figure}

\subsection{Steering Control Performance of Non-PD Participants}
\begin{figure}[H]
    \centering
    \includegraphics[width=0.7\linewidth]{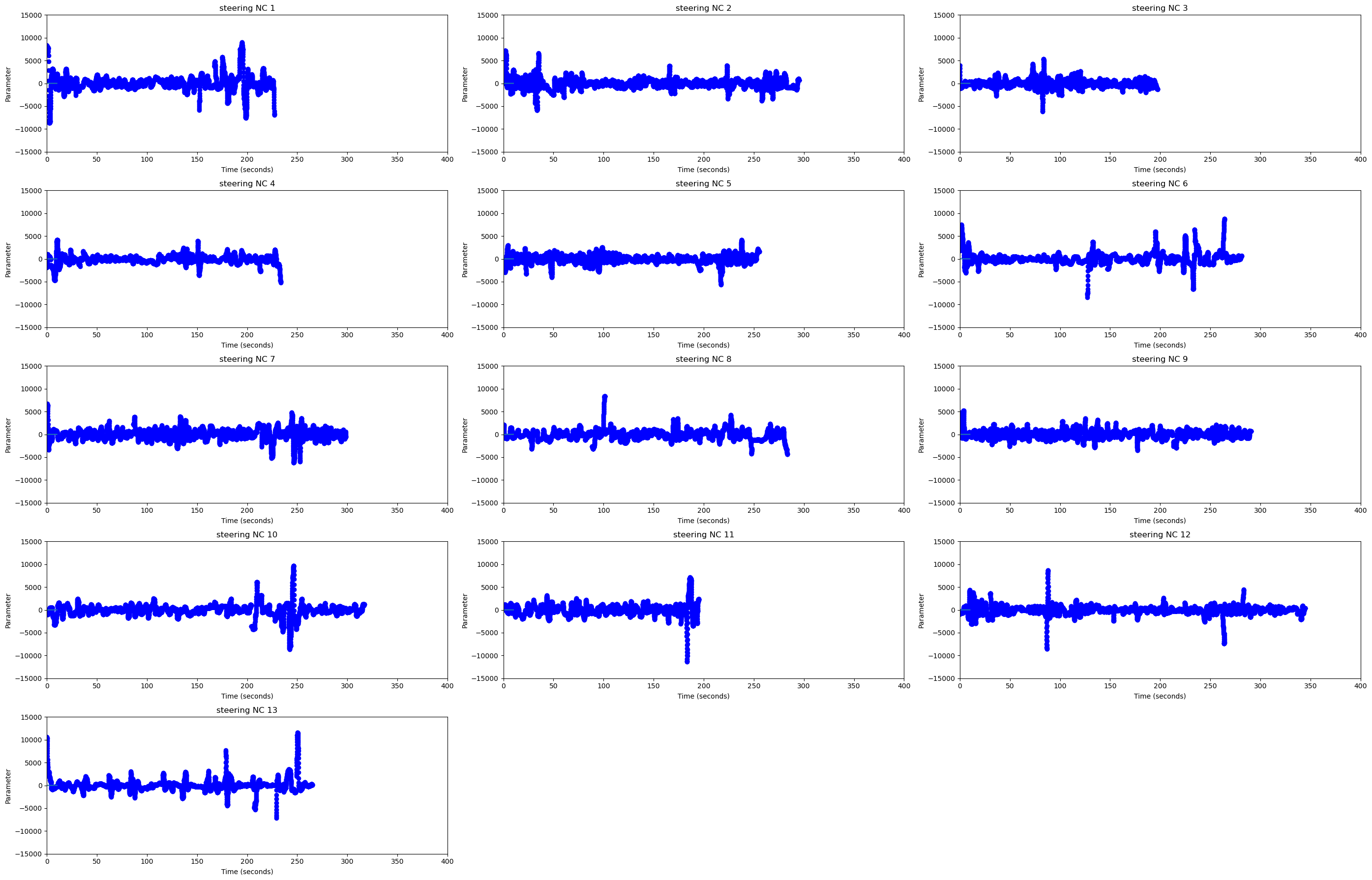}
    \caption{This figure sequentially presents the steering control performance of NC 1 to NC 13 during the data collection process. In each chart, the horizontal axis represents time in seconds, while the vertical axis represents the steering wheel angle. The zero point on the vertical axis (center of the y-axis) indicates a steering wheel angle of 0, with positive values representing a right turn and negative values representing a left turn.}
    \Description{This figure consists of 13 line plots arranged in a grid, each showing the steering control performance of an NC participant during the data collection process. The participants are labeled NC 1 to NC 13, with each plot representing their steering data. The x-axis in all plots represents time in seconds (ranging from 0 to 400), and the y-axis represents a parameter related to the steering angle, with values ranging from -15,000 to 15,000.}
    \label{fig:enter-label1}
\end{figure}

\subsection{Throttle Control Performance of People with PD}
\begin{figure}[H]
    \centering
    \includegraphics[width=0.8\linewidth]{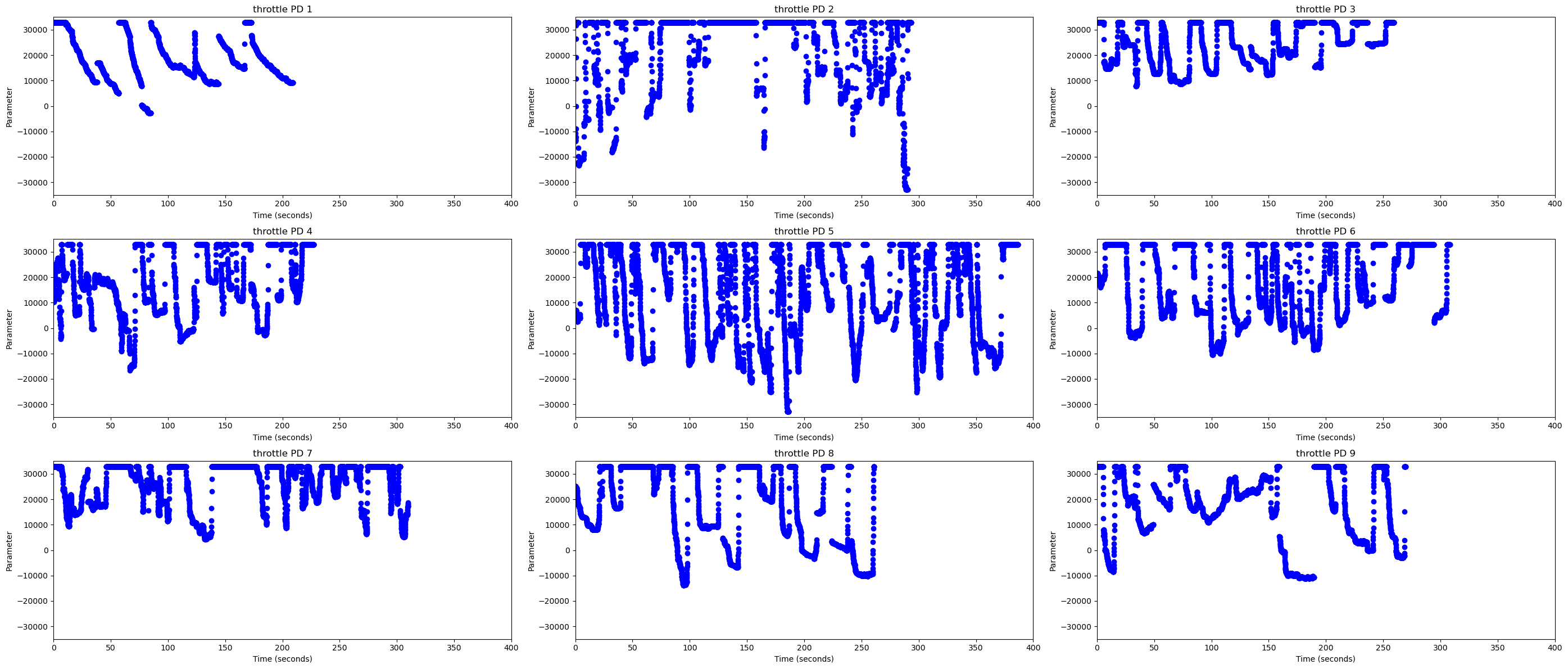}
    \caption{This figure sequentially presents the throttle control performance of PD 1 to PD 9 during the data collection process. In each chart, the horizontal axis represents time in seconds, while the vertical axis represents the pedal depth. The highest values (at the top of the y-axis) represent fully released pedals, while lower values indicate deeper pedal engagement.}
    \Description{This figure consists of nine line plots arranged in a 3 × 3 grid. Each plot represents the throttle control behavior of a person with PD over time, labeled PD 1 through PD 9. The x-axis in all plots represents time in seconds (ranging from 0 to 400), and the y-axis represents a parameter related to the pedal depth, with values ranging from -35,000 to 35,000.}
    \label{fig:enter-label2}
\end{figure}

\subsection{Throttle Control Performance of Non-PD Participants}
\begin{figure}[H]
    \centering
    \includegraphics[width=0.8\linewidth]{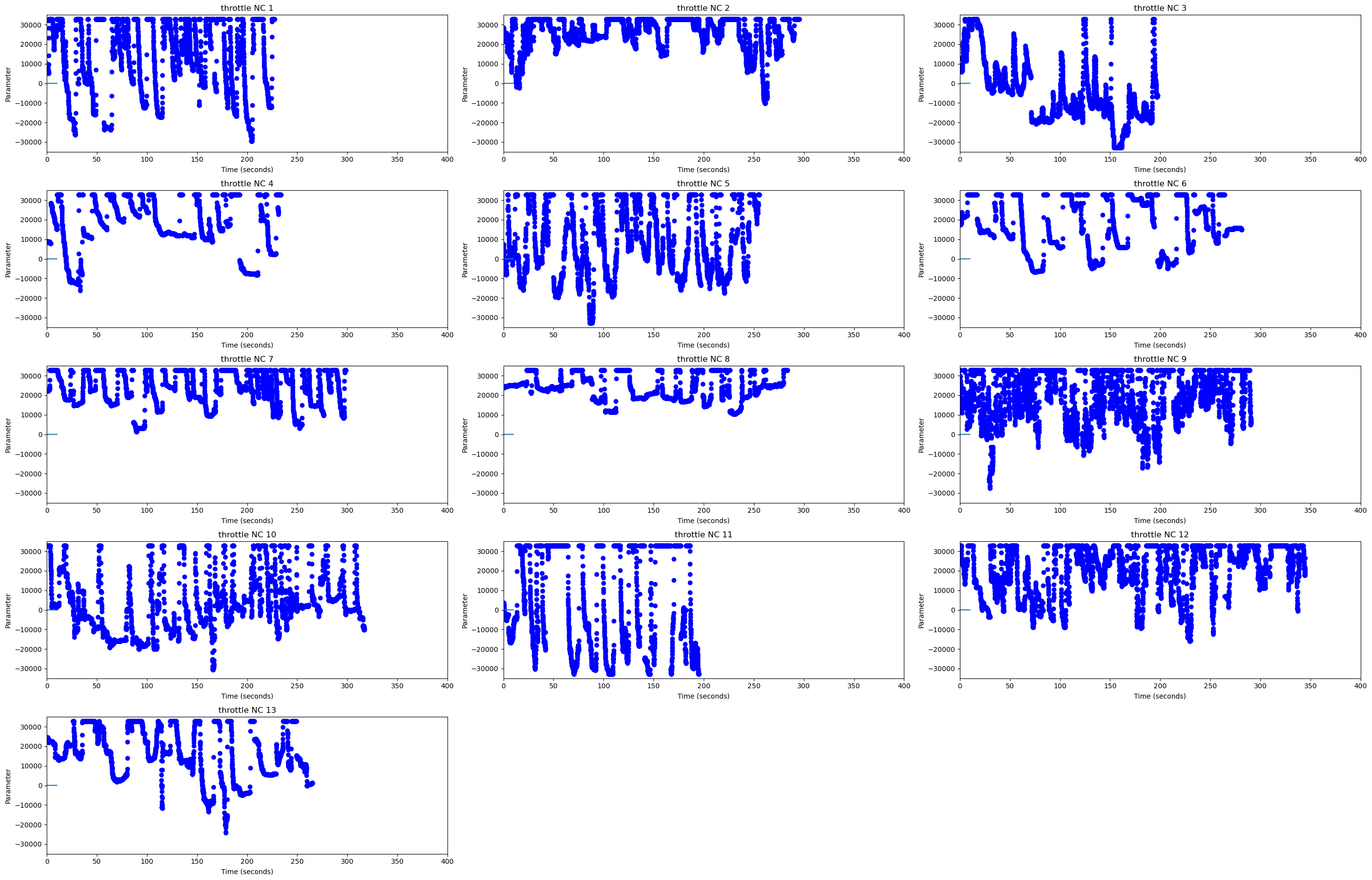}
    \caption{This figure sequentially presents the throttle control performance of NC 1 to NC 13 during the data collection process. In each chart, the horizontal axis represents time in seconds, while the vertical axis represents the pedal depth. The highest values (at the top of the y-axis) represent fully released pedals, while lower values indicate deeper pedal engagement.}
    \Description{This figure consists of 13 line plots arranged in a grid, each showing the throttle control performance of an NC participant during the data collection process. The participants are labeled NC 1 to NC 13, with each plot representing their throttle data. The x-axis in all plots represents time in seconds (ranging from 0 to 400), and the y-axis represents a parameter related to the pedal depth, with values ranging from -35,000 to 35,000.}
    \label{fig:enter-label3}
\end{figure}

\subsection{Brake Control Performance of People with PD}
\begin{figure}[H]
    \centering
    \includegraphics[width=0.8\linewidth]{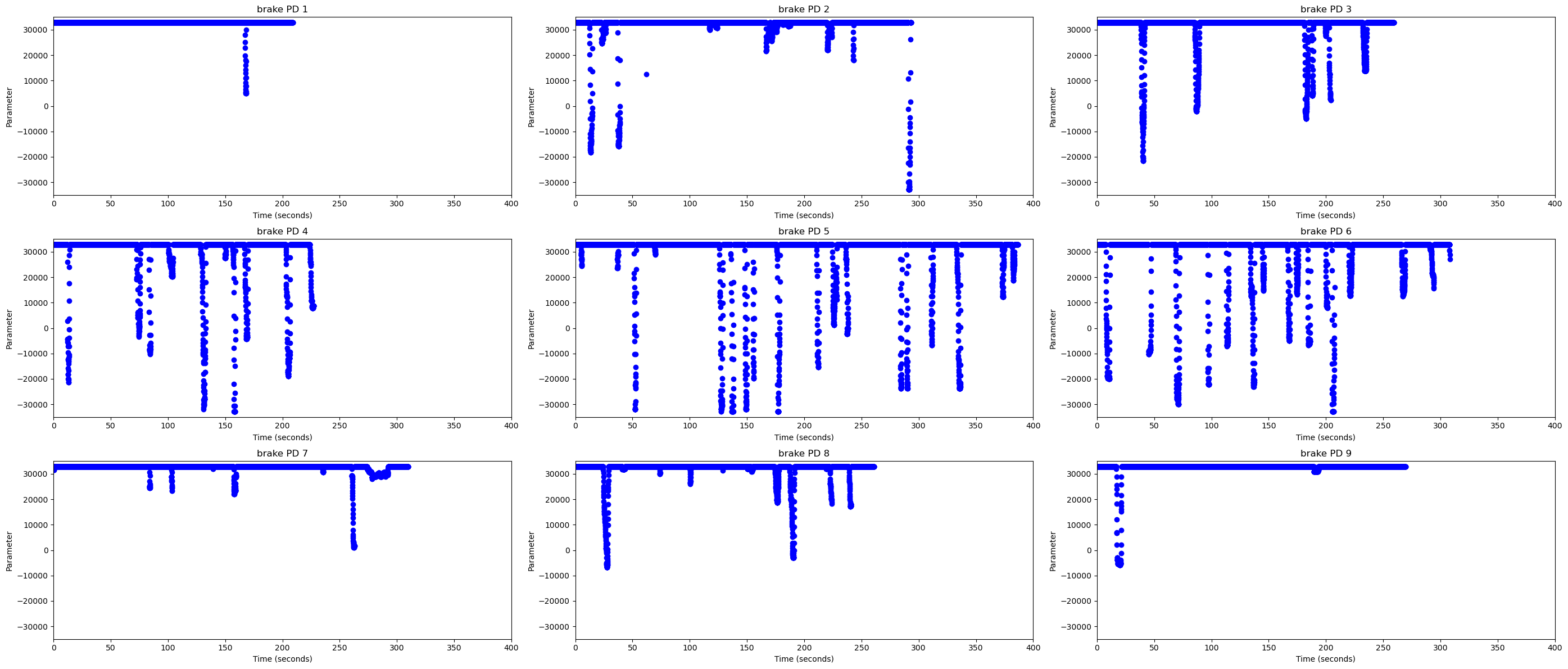}
    \caption{This figure sequentially presents the brake control performance of PD 1 to PD 9 during the data collection process. In each chart, the horizontal axis represents time in seconds, while the vertical axis represents the pedal depth. The highest values (at the top of the y-axis) represent fully released pedals, while lower values indicate deeper pedal engagement.}
    \Description{This figure consists of nine line plots arranged in a 3 × 3 grid. Each plot represents the brake control behavior of a person with PD over time, labeled as PD 1 through PD 9. The x-axis in all plots represents time in seconds (ranging from 0 to 400), and the y-axis represents a parameter related to the pedal depth, with values ranging from -35,000 to 35,000.}
    \label{fig:enter-label4}
\end{figure}

\subsection{Brake Control Performance of Non-PD Participants}
\begin{figure}[H]
    \centering
    \includegraphics[width=0.8\linewidth]{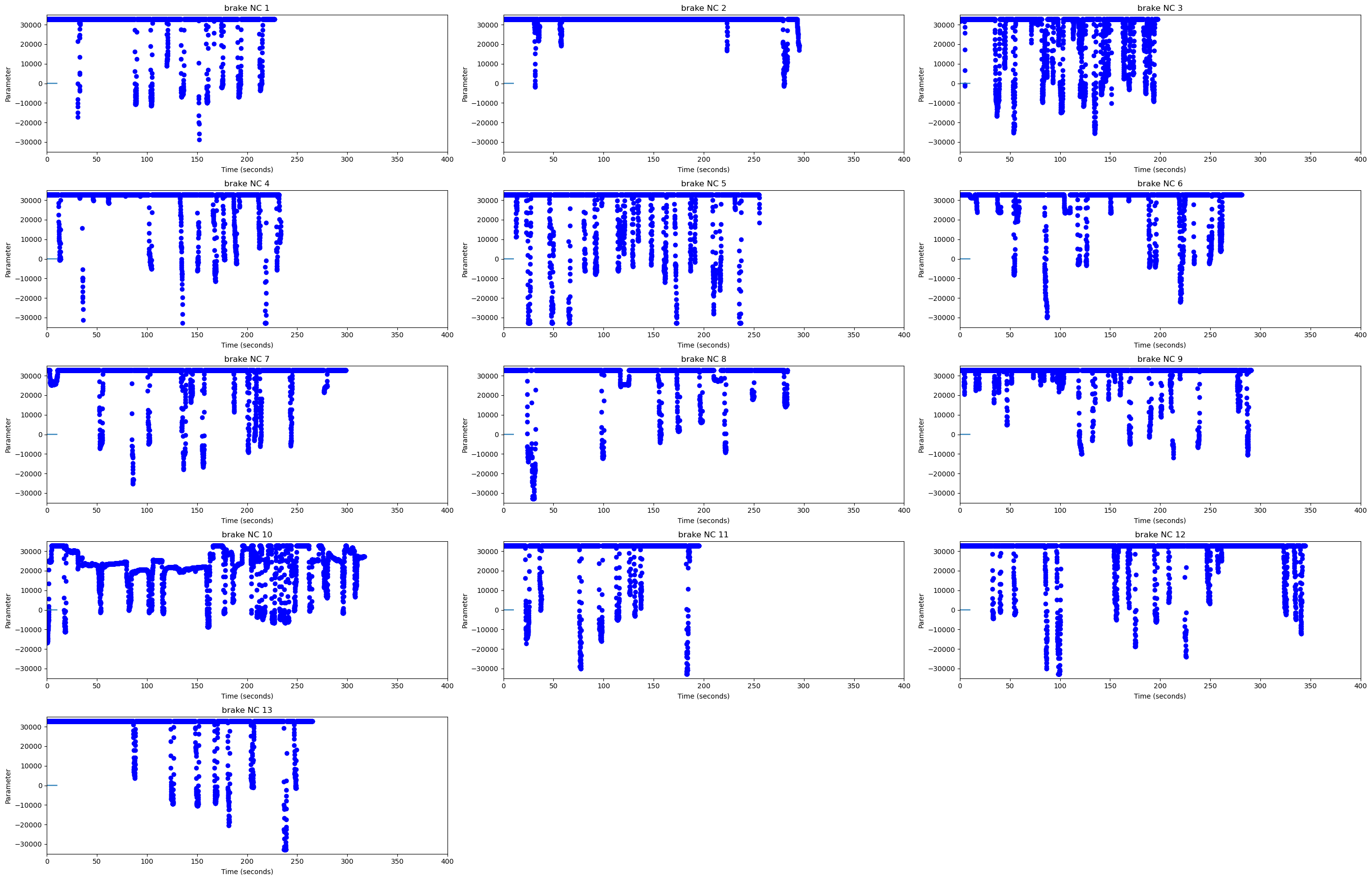}
    \caption{This figure sequentially presents the brake control performance of NC 1 to NC 13 during the data collection process. In each chart, the horizontal axis represents time in seconds, while the vertical axis represents the pedal depth. The highest values (at the top of the y-axis) represent fully released pedals, while lower values indicate deeper pedal engagement.}
    \Description{This figure consists of 13 line plots arranged in a grid, each showing the brake control performance of an NC participant during the data collection process. The participants are labeled NC 1 to NC 13, with each plot representing their brake data. The x-axis in all plots represents time in seconds (ranging from 0 to 400), and the y-axis represents a parameter related to the pedal depth, with values ranging from -35,000 to 35,000.}
    \label{fig:enter-label5}
\end{figure}
\clearpage

\section{Thematic Analysis Tables}
In this section, we present thematic analysis results from the preliminary study and the user study.





\subsection{Thematic Analysis Results in the Preliminary Study with People with PD}
\begin{table}[h]
\centering
\begin{tabular}{ccp{6cm}cc}

\hline
\textbf{Theme} & \textbf{Subtheme} & \centering\textbf{Theme description} & \textbf{Number of participants} & \textbf{\%}\\
\hline
\multirow{8}*{Demand} & \multirow{2}*{Live independently} & 
\centering People with PD demonstrated a desire to live independently & \multirow{2}*{8} & \multirow{2}*{72.7}\\
~ & \multirow{4}*{Driving} & \centering Maintaining the ability to drive is crucial for preserving independence and plays a significant role in the
self-esteem of people with PD & \multirow{4}*{8} & \multirow{4}*{72.7}\\
~ & \multirow{2}*{Access driving ability} & \centering People with PD have a desire to access their driving ability & \multirow{2}*{4} & \multirow{2}*{36.4}\\

\hline
\multirow{6}*{Factors Impact Driving} & \multirow{2}*{"On-off" effect of PD} & \centering Participants noted that the driving state is influenced by the fluctuation of PD symptoms & \multirow{2}*{9} & \multirow{2}*{81.8}\\
~ & \multirow{2}*{Sleep attacks} & \centering Participants reported frequently feeling drowsy or sleepy while driving & \multirow{2}*{6} & \multirow{2}*{54.5}\\
~ & \multirow{2}*{Weather influence} & \centering Participants reported difficulties when driving in strong sunlight & \multirow{2}*{3} & \multirow{2}*{27.3}\\

\hline
\multirow{6}*{Challenges} & \multirow{2}*{Reaction speed} & \centering Participants noted that their reactions had slowed after developing PD & \multirow{2}*{9} & \multirow{2}*{81.8}\\
~ & \multirow{2}*{Multitasking ability} & \centering Participants noted that their multitasking ability decreased after developing PD & \multirow{2}*{4} & \multirow{2}*{36.4}\\
~ & \multirow{2}*{Privacy} & \centering Designing an alert system for
people with PD must address privacy and social considerations & \multirow{2}*{6} & \multirow{2}*{54.5}\\

\hline
\end{tabular}
\captionsetup{position=bottom} 
\caption{The thematic analysis result in the preliminary study for interviews with 11 people with PD.} 
\Description{
    This table summarizes key themes, subthemes, descriptions, and participant feedback from a study involving people with PD. It includes three main themes: "Demand," "Factors Impact Driving," and "Challenges."
    
    Demand: Highlights the desire of people with PD to live independently (72.7\%), maintain the ability to drive (72.7\%), and understand their driving ability (36.4\%).
    
    Factors Impact Driving: Focuses on elements affecting driving performance, such as the "on-off" effect of PD symptoms (81.8\%), sleep attacks or drowsiness while driving (54.5\%), and difficulties caused by weather, like strong sunlight (27.3\%).
    
    Challenges: Addresses issues like slowed reaction speed (81.8\%), reduced multitasking ability (36.4\%), and the importance of considering privacy and social factors when designing an alert system (54.5\%).
    
    Each subtheme includes the percentage of participants who mentioned it, providing insights into the challenges and needs of people with PD.
}
\label{tab:thematic analysis pre patients}
\end{table}

\subsection{Thematic Analysis Results in the Preliminary Study with PD Specialists}




\begin{table}[H]
\centering
\begin{tabular}{ccp{6cm}cc}
\hline
\textbf{Theme} & \textbf{Subtheme} & \centering \textbf{Theme description} & \textbf{Number of participants} & \textbf{\%}\\
\hline
\multirow{5}*{Current Solutions} & \multirow{3}*{Outpatient concerns} & \centering PD specialists in outpatient clinics do not typically focus on whether patients can maintain their driving ability & \multirow{3}*{4} & \multirow{3}*{100}\\
~ & \multirow{2}*{Guidelines} & \centering No uniform legal criteria to guide PD specialists on assessing driving fitness & \multirow{2}*{4} & \multirow{2}*{100}\\

\hline
\multirow{4}*{Factors Impact Driving} & \multirow{2}*{"On-off" effect of PD} & \centering PD specialist mentioned the "on-off" phenomenon of PD affects driving & \multirow{2}*{3} & \multirow{2}*{75}\\
~ & \multirow{2}*{Sleep attacks} & 
\centering PD specialist mentioned the impact of drowsiness on driving & \multirow{2}*{3} & \multirow{2}*{75}\\

\hline
\multirow{6}*{Challenges} & \multirow{2}*{Reaction speed} & \centering PD specialist noted people with PD have a slower reaction speed & \multirow{2}*{4} & \multirow{2}*{100}\\
~ & \multirow{2}*{Multitasking ability} & \centering PD specialist noted people with PD have impaired multitasking ability & \multirow{2}*{2} & \multirow{2}*{50}\\
~ & \multirow{2}*{Privacy} & \centering Designing an alert system for
people with PD must address privacy and social considerations & \multirow{2}*{2} & \multirow{2}*{50}\\

\hline
\end{tabular}
\captionsetup{position=bottom} 
\caption{The thematic analysis result in the preliminary study for interviews with 4 PD specialists.} 
\Description{
    This table provides insights into themes discussed by PD specialists, including current solutions, factors impacting driving, and challenges faced by people with PD. 
    Current Solutions:  
    PD specialists noted that outpatient clinics do not typically assess patients' ability to drive (100\%).
    There are no uniform legal guidelines to help PD specialists evaluate driving fitness (100\%).
    Factors Impacting Driving:
    The "on-off" phenomenon of PD symptoms significantly affects driving ability (75\%).
    Drowsiness, or sleep attacks, also impacts driving performance (75\%).
    Challenges: 
    Slower reaction speed is a universal issue for people with PD (100\%).
    Multitasking ability is often impaired (50\%).
    Privacy and social considerations are important factors when designing an alert system (50\%).
    The table highlights the perspectives of PD specialists on these critical topics, along with the mention rate among participants.
}
\label{tab:thematic analysis pre specialists}
\end{table}

\subsection{Thematic Analysis Results in the User Study}




\begin{table}[H]
\centering
\begin{tabular}{ccp{6cm}cc}
\hline
\textbf{Theme} & \textbf{Subtheme} & \centering\textbf{Theme description} & \textbf{Number of participants} & \textbf{\%}\\
\hline
\multirow{5}*{Feedback of the System} & \multirow{2}*{Acceptance level} & \centering Participants recognized the significance of our system & \multirow{2}*{5} & \multirow{2}*{100}\\
~ & \multirow{3}*{Useful scenarios} & \centering People with PD thought this system is useful when they experience fluctuations in their driving performance & \multirow{3}*{4} & \multirow{3}*{80}\\

\hline
\multirow{4}*{Acceptance of Scenarios} & \multirow{2}*{Acceptable scenarios} & \centering Scenarios where independent driving is more acceptable & \multirow{2}*{3} & \multirow{2}*{60}\\
~ & \multirow{2}*{Unacceptable scenarios} & \centering Driving with other passengers in the car is less acceptable due to privacy concerns & \multirow{2}*{3} & \multirow{2}*{60}\\

\hline
\end{tabular}
\captionsetup{position=bottom} 
\caption{The thematic analysis result in the user study with 5 people with PD.} 
\Description{
    This table summarizes participant feedback on the system and their acceptance of various scenarios. It highlights two main themes: "Feedback of the System" and "Acceptance of Scenarios."    
    Feedback of the System:
    Acceptance level: All participants (100\%) recognized the significance of the system.
    Useful scenarios: 80\% of participants mentioned the system is useful when they experience fluctuations in driving performance.
    Acceptance of Scenarios:
    Acceptable scenarios: 60\% of participants found independent driving scenarios to be more acceptable.
    Unacceptable scenarios: 60\% of participants considered driving with other passengers less acceptable due to privacy concerns.
    This table provides an overview of how participants view the system and its applicability in various driving situations.
}
\label{tab:thematic analysis user study}
\end{table}

\end{document}